\documentclass[floatfix,aps,prb,twocolumn,superscriptaddress]{revtex4-2}
\usepackage{float}
\restylefloat{table}
\usepackage{adjustbox}
\usepackage{tikz}
\usetikzlibrary{calc,positioning}
\usepackage{amsmath,amssymb}
\usepackage{subfig}
\usepackage{graphicx}
\usepackage{braket}
\usepackage{epstopdf}
\usepackage{hyperref}
\usepackage{float}
\usepackage{xcolor}
\usepackage{tikz}
\usetikzlibrary{positioning}
\usepackage{graphicx}
\usepackage{dcolumn}
\usepackage{bm}
\usepackage[normalem]{ulem}
\usepackage{amsmath,amssymb}
\usepackage{subfig}
\usepackage{graphicx}
\usepackage{braket}
\usepackage{epstopdf}
\usepackage{hyperref}
\usepackage{float}
\usepackage{xcolor}
\usepackage{tikz}
\usetikzlibrary{positioning}
\usepackage{amsmath,mleftright}
\usepackage{xparse}
\usepackage{ragged2e}
\usepackage{caption}
\usepackage{amsmath}
\usepackage{hyperref}
\usepackage{graphicx}
\usepackage{float}
\usepackage{soul}
\usepackage{color}
\usepackage{braket}
\usepackage{tikz}
\newcommand{\comment}[1]{}

\newcommand{\lt} {\left}
\newcommand{\rt} {\right}
\newcommand{\up} {\uparrow}
\newcommand{\dn} {\downarrow}

\begin{document}
\title{
Sign changes of the thermoelectric transport coefficient \\
across the metal insulator crossover in the doped Fermi Hubbard model} 



\author{Sayantan Roy}
\affiliation{Department of Physics, The Ohio State University, Columbus OH 43210, USA}
\author{Abhisek Samanta}
\affiliation{Department of Physics, The Ohio State University, Columbus OH 43210, USA}
\affiliation{Department of Physics, Indian Institute of Technology
	Gandhinagar, Gujarat 382355, India}
\author{Nandini Trivedi}
\affiliation{Department of Physics, The Ohio State University, Columbus OH 43210, USA}



\date{\today}

\begin{abstract}
 
We investigate the doping-dependence of the Seebeck coefficient, as calculated from the Kelvin formula, for the Fermi Hubbard model using determinantal quantum Monte Carlo simulations. Our key findings are: (1) Besides the expected hole to electron-like behavior change around half filling, we show that the additional sign change at an electronic density $n_s$ (and correspondingly a hole density $p_s$) is controlled by the opening of a charge gap in the {\em thermodynamic} density of states or compressibility and not by the pseudogap scale in the single particle density of states. 
  (2) We find that $n_s(T,U)$ depends strongly on the interaction $U$ and shows an unusual non-monotonic dependence on temperature with a maximum at a temperature $T\approx t$, on the order of the hopping scale. (3) We identify local moment formation close to half filling as the main driver for the anomalous behavior of the thermoelectric transport coefficient. 
  
\end{abstract}


\maketitle


\textit{Introduction:} Of particular interest in strongly correlated systems is the idea of emergence~\cite{anderson1972more}, where collective behavior with markedly novel properties emerge due to interaction effects.
When a Mott insulator (which has an odd number of electrons per unit cell, with strong interactions) is slightly doped, 
the system can behave very differently 
from a Landau Fermi liquid with well defined quasiparticles \cite{varma2016quantum,zaanen2019planckian,sachdev2023quantum}.
Experimental detection of such novel phases remains a considerable effort in understanding strongly correlated systems \cite{mitrano2018anomalous,chen2019incoherent,ekahana2024anomalous,haslinger2002arpes,hussey2008phenomenology,damascelli2003angle,shen1999novel,sarkar2019correlation,proust2019remarkable,krockenberger2012unconventional,abdel2006anisotropic,badoux2016change}.

Transport is one of the first probes of quantum materials and can reveal striking puzzles about nature of excitations in these enigmatic phases. 
Thermopower or the Seebeck coefficient is an important transport quantity which measures the efficiency of direct conversion from thermal to electrical energy. In addition, it tracks the nature of carriers in the system. In weakly interacting systems, where Fermi liquid theory holds, the Seebeck coefficient is positive when the excitations are electron-like, but changes to negative when excitations are hole-like~\cite{behnia2015fundamentals}.
However, the presence of strong correlations in the system can lead to \textit{anomalous} behavior, including a change in sign and magnitude of the Seebeck coefficient compared to Fermi liquid predictions. Earlier experiments found universal zero crossings of the Seebeck coefficient near the optimal doping of cuprates for a large class of materials~\cite{obertelli1992systematics,cooper1987thermoelectric}. More recent experiments on thermopower have offered a wealth of information; from sign changes in the Seebeck coefficient attributed to the Fermi surface reconstruction~\cite{gourgout2022seebeck,laliberte2011fermi,cyr2017anisotropy,badoux2016critical,chang2010nernst}, signatures of quantum fluctuations near critical points \cite{mandal2019anomalous,wakamatsu2023thermoelectric,izawa2007thermoelectric}, violation of Fermi liquid behavior in twisted bilayer graphene \cite{ghawri2022breakdown,paul2022interaction},
to novel flatband physics \cite{zhang2022exchange,garmroudi2024high,chen2022anomalous}.


\begin{figure*}
\begin{tikzpicture}

\node (img1) {\includegraphics[width=0.29\linewidth]{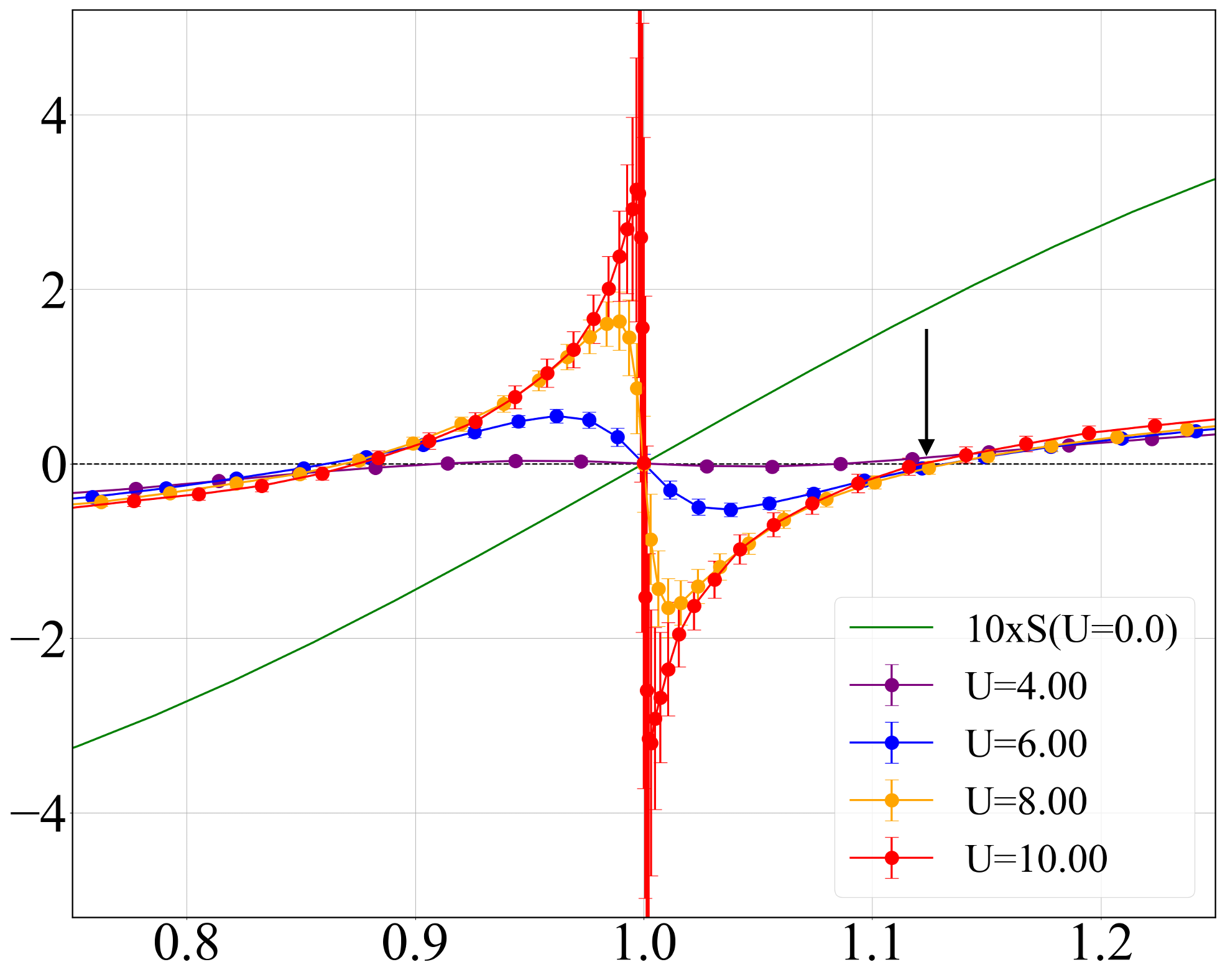}};
\node[left=of img1,node distance=0cm,rotate=90,anchor=center,yshift=-1.0cm,xshift=0.0cm]{\small{ S$_{\text{kelvin}}[k_B/e]$}};
\node[below=of img1,node distance=0cm,yshift=1.05cm,xshift=0.0cm]{\large{$n$}};
\node[left=of img1,node distance=0cm,yshift=1.7cm,xshift=2.2cm]{\normalsize{(a)}};
\node[left=of img1,node distance=0cm,yshift=1.7cm,xshift=6.2cm]{\normalsize{{T = 0.5}}};
\node[left=of img1,node distance=0cm,yshift=0.5cm,xshift=5.5cm]{\tiny{{$n_s$}}};
\node (img2) [right=of img1,xshift=-0.55cm]{\includegraphics[width=0.29\linewidth]{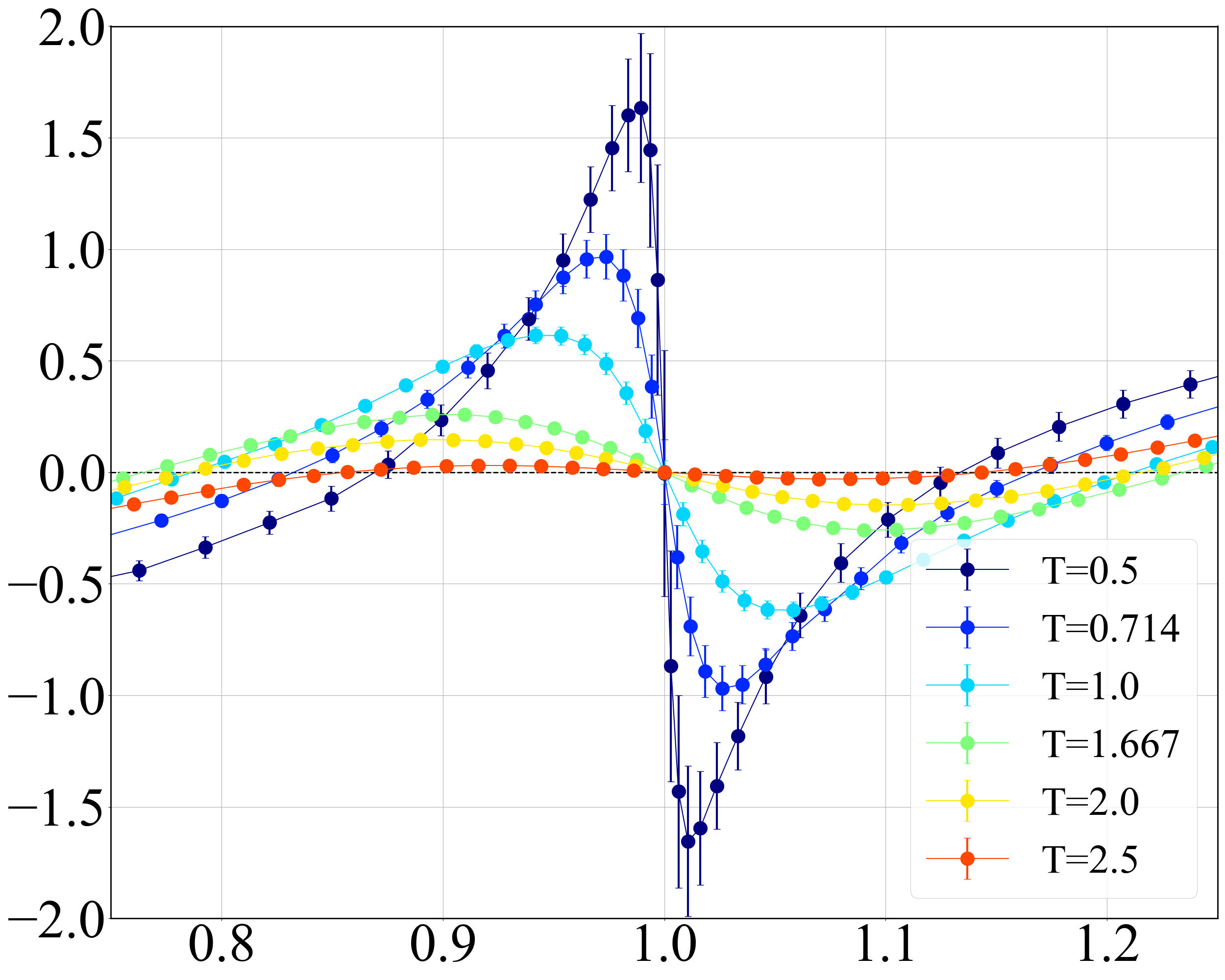}};
\node[left=of img2,node distance=0cm,rotate=90,anchor=center,yshift=-1.0cm,xshift=0.0cm]{\small{ S$_{\text{kelvin}}[k_B/e]$}};
\node[left=of img2,node distance=0cm,yshift=1.7cm,xshift=6.2cm]{\normalsize{{U = 8.0}}};
\node[below=of img2,node distance=0cm,yshift=1.0cm,xshift=0.0cm]{\large{$n$}};
\node[left=of img2,node distance=0cm,yshift=1.7cm,xshift=2.4cm]{\normalsize{(b)}};
\node (img3) [right=of img2,xshift=-0.55cm]{\includegraphics[width=0.27\linewidth]{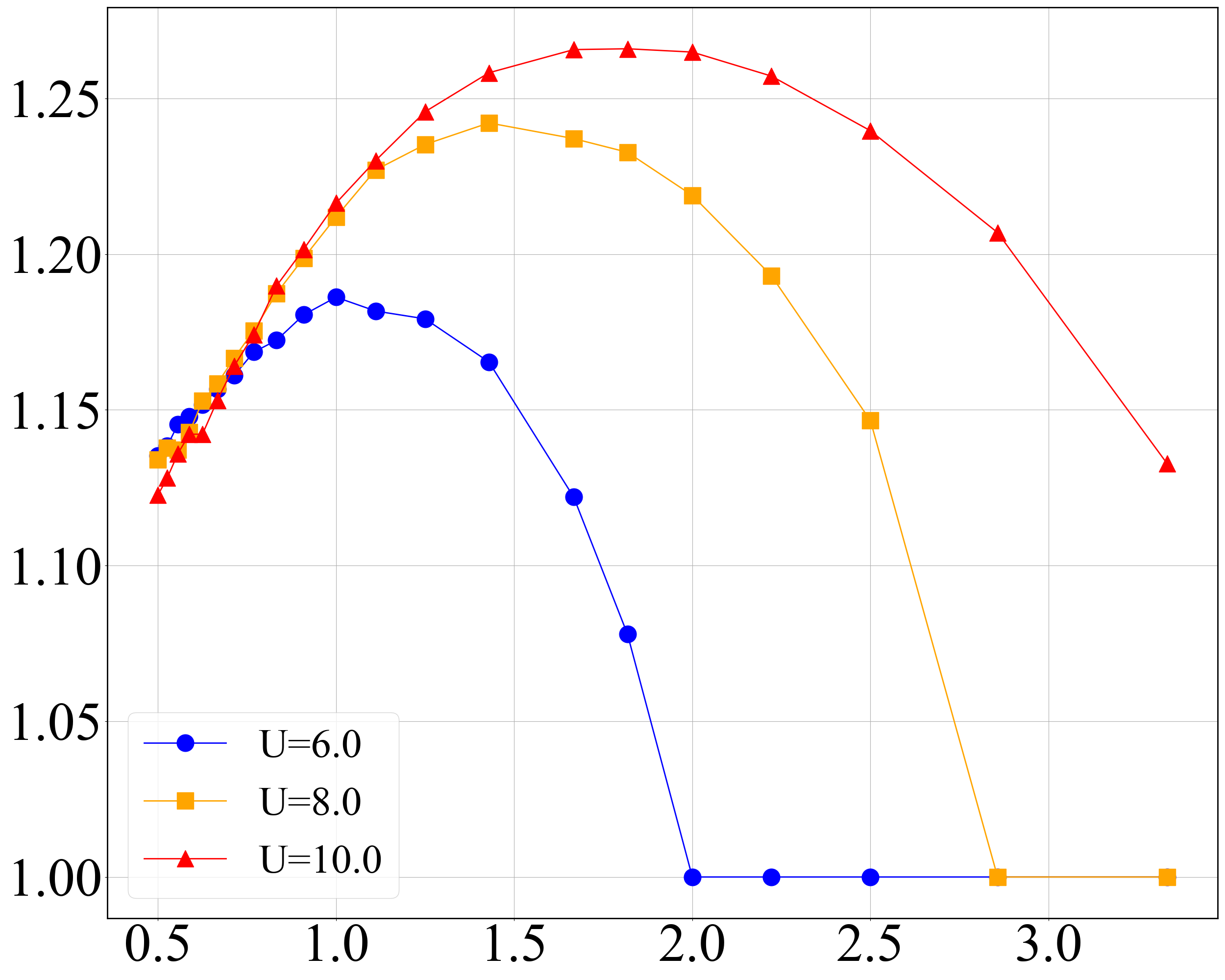}};
\node[left=of img3,node distance=0cm,yshift=0.0cm,xshift=1.3cm]{\large{$n_s$}};
\node[below=of img3,node distance=0cm,yshift=1.0cm,xshift=0.0cm]{\large{$T$}};
\node[left=of img3,node distance=0cm,yshift=1.6cm,xshift=2.2cm]{\normalsize{(c)}};
\end{tikzpicture}
\captionof{figure}{Seebeck coefficient in the repulsive Hubbard model on a square lattice at intermediate to high temperatures. \textbf{(a)} Seebeck coefficient $S_{\rm kelvin}$, as calculated from the Kelvin formula (defined in text) at a temperature of $T = 0.5$. Note that the anomalous sign change develops close to $U = 3.2$ at this temperature.\textbf{(b)} Seebeck coefficient calculated at $U = 8.0$ for multiple $T$. Note that with increasing temperature, the Seebeck anomaly (wrong sign of the Seebeck coefficient) happens over a larger doping window. \textbf{(c)} Doping $n_s$ at which the Seebeck coefficient changes sign to match the ``correct" carrier type, vs temperature for multiple $U$. Note that the free particle limit is not recovered monotonically; $n_s$ increases up to $T\sim O(t)$ and then decreases to half filling at a temperature scale set by $U$.} 
\label{Seebeck_coefficient}
\end{figure*}
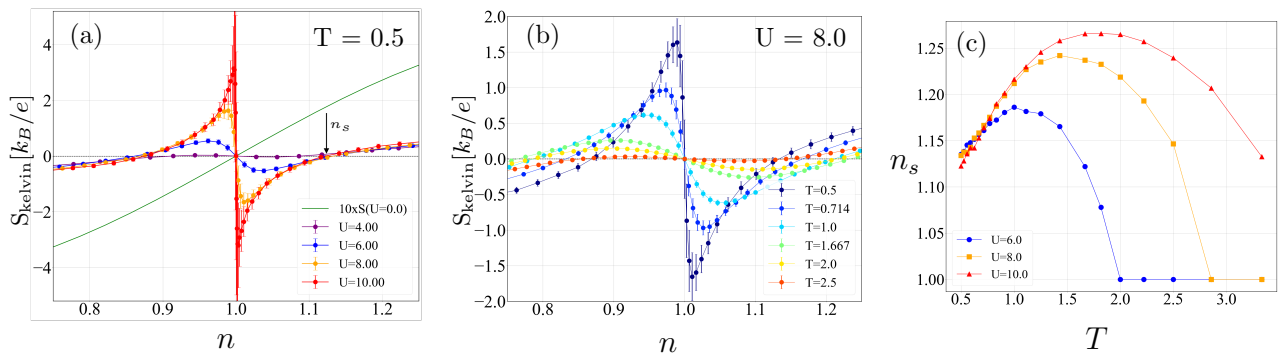

Numerical simulations of transport quantities serve as an important benchmark to understand experimental observations. Over the last decade, several calculations of the Seebeck coefficient have been reported for both model Hamiltonians \cite{silva2023effects,arsenault2013entropy,georges2021skewed,mravlje2016thermopower,davison2017thermoelectric,wang2023quantitative,xu2013hidden} and realistic materials \cite{suzuki2023distinct,tomczak2010thermopower,chikina2020correlated,mukerjee2007doping,grissonnanche2022seebeck,mravlje2016thermopower,wang2018first,yang2016density,takaki2017first}.

These calculations have established a host of phenomena, e.g. effect of particle hole symmetry on the doping dependence of Seebeck coefficient \cite{silva2023effects}, the relevance of an effective Hubbard-like model for describing cuprate physics~\cite{wang2023quantitative}, 
skew scattering influencing the anomalous sign change of Seebeck coefficient in non-Fermi liquids~\cite{georges2021skewed}, and emergence of a ``Hund's metal" like behavior when multiple orbital degrees of freedom contribute to transport \cite{mravlje2016thermopower,suzuki2023distinct}. Seebeck coefficient has provided information about quantum criticality in high-$T_c$ superconductors~\cite{garg2011thermopower} and heavy fermion systems \cite{kim2010thermopower}, 
in addition to identifying a crossover region between Fermi liquid behavior and Mott-Ioffe-Regel limit with resilient quasiparticles~\cite{deng2013bad}. 

Several questions remain to be answered and form the focus of this Letter: 
i) What drives the anomalous sign change of the Seebeck coefficient in the vicinity of Mott insulating phase? 
ii) Can one find a footprint of the Seebeck anomaly in other thermodynamic observables? If so, what are the temperature and interaction dependencies of such behavior?
iii) Since strong correlations tend to localize electrons and form local moments, what role do they play in the Seebeck anomaly? What are the probes to test such behavior?


We derive insights about the Seebeck coefficient $S$  in the strongly correlated regime using the Kelvin formula $S_{\rm Kelvin} = -\frac{1}{e}\big(\frac{\partial s}{\partial n}\big)_{T}$ \cite{peterson2010kelvin} represented in terms of the entropy density. This can be obtained from the Kubo formula in the slow (thermodynamic) limit instead of the fast (transport) limit while evaluating the Onsager coefficients. Our main results are the following:

\noindent (1) At intermediate to high temperatures, the appearance of anomalous zero crossings $n_s$ of the Seebeck coefficient $S_{\rm Kelvin}$ is governed by the opening of a charge gap in the thermodynamic density of states (TDOS). 

\noindent (2) In presence of strong correlations, $n_s(T)$ shows a non-monotonic behavior finally approaching the expected half filling point at a temperature set by $U$.

\noindent (3) The anomalous phase where Seebeck coefficient has the opposite sign compared to Fermi liquid theory is primarily dominated by the formation of local moments. 


\textit{Model and method:} We consider the single band Fermi Hubbard model on a square lattice with nearest neighbor hopping and onsite repulsive interaction,
$\mathcal{H}= -t\sum_{\langle ij \rangle,\sigma}{  \hat{c}^\dag_{i\sigma} \hat{c}_{j\sigma} }-\mu \sum_{i}\hat{n}_{i}+U\sum_{i}{ \left( \hat{n}_{i\uparrow} - \frac{1}{2} \right)\left( \hat{n}_{i\downarrow} - \frac{1}{2} \right)}
$. The operators
 $\hat{c}_{i\sigma}$ and $\hat{c}^\dag_{i\sigma}$ are fermionic annihilation and creation operators, respectively.
The number operator is defined as $\hat{n}_{i,\sigma} \equiv \hat{c}^\dag_{i\sigma} \hat{c}_{i\sigma}$,  $\hat{n}_i = \hat{n}_{i \uparrow} + \hat{n}_{i\downarrow}$, and the particle density per site ${n} = \sum_i{\langle\hat{n}_i\rangle}/{N_s}$, where $N_s$ is the total number of sites. We define hopping amplitude $t$ as the energy scale, $\mu$ is the chemical potential and $U$ is the onsite Coulomb repulsion. We perform numerically exact Determinantal Quantum Monte Carlo (DQMC) ~\cite{blankenbecler1981monte,hirsch1985two} at intermediate to high interaction strengths. 
We perform analytic continuation using the maximum entropy package CQMP-MaxEnt~\cite{levy2017implementation}, with default models chosen to optimize the sum rules~\cite{white1991spectral,swanson2014dynamical}. We also construct a semi-analytic parton mean field theory in Appendix~\ref{Parton MFT} to capture the effect of charge gap on the Seebeck anomaly.

The evolution of the Seebeck coefficient $S_{\rm Kelvin}(n,U,T)$ as a function of density $n$, interaction strength $U$ and temperature $T$, is shown in Fig.~\ref{Seebeck_coefficient}(a)-(b). With increasing interaction strength, there is a deviation of the Seebeck coefficient from Fermi liquid like behavior and appearance of an anomalous zero crossing at $n_s(T,U)$ with a ``wrong" sign of the Seebeck coefficient that develops near the Mott insulator at half filling. 
$n_s (T,U)$ shows a striking behavior with temperature in Fig.~\ref{Seebeck_coefficient}(c). Contrary to Hall coefficient calculations, in which the anomaly vanishes monotonically with increasing temperature~\cite{khait2023hall}, the Seebeck anomaly {\em increases} with increasing temperature, saturates at a temperature scale set by the interaction strength before eventually approaching the free particle behavior. As we discuss below, this anomalous behavior can be understood from the entropic origin of thermopower, and the formation of local moments that dominate transport in this regime. 

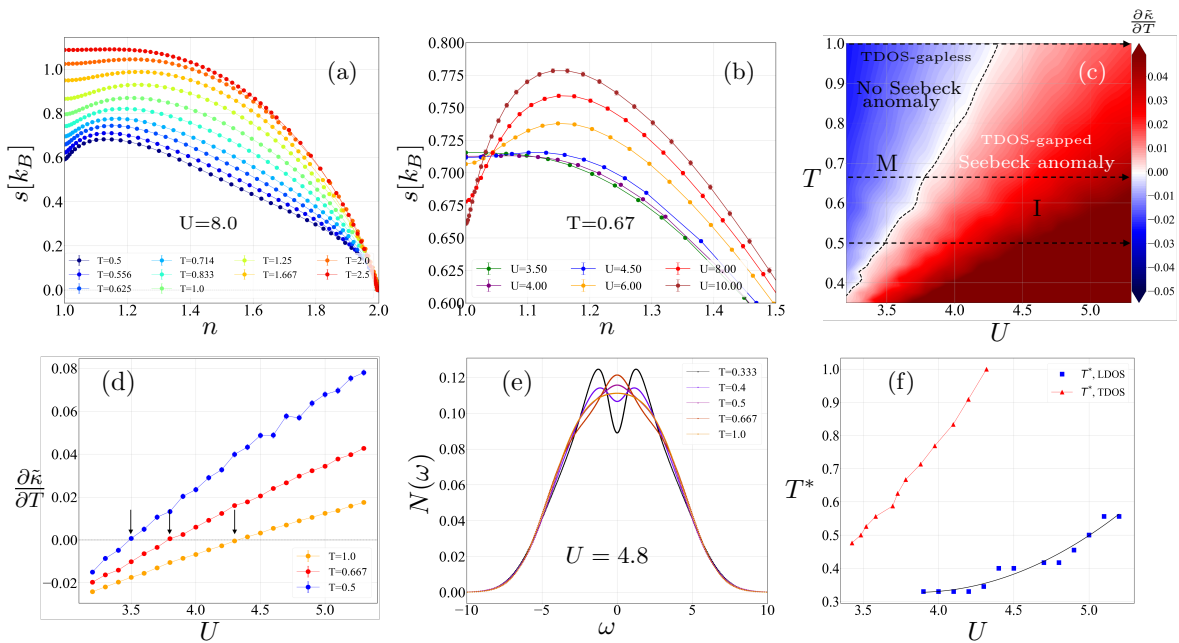
\begin{figure*}[t]
\begin{tikzpicture}
\node (img1) {\includegraphics[width=4.8cm,height=3.8cm]{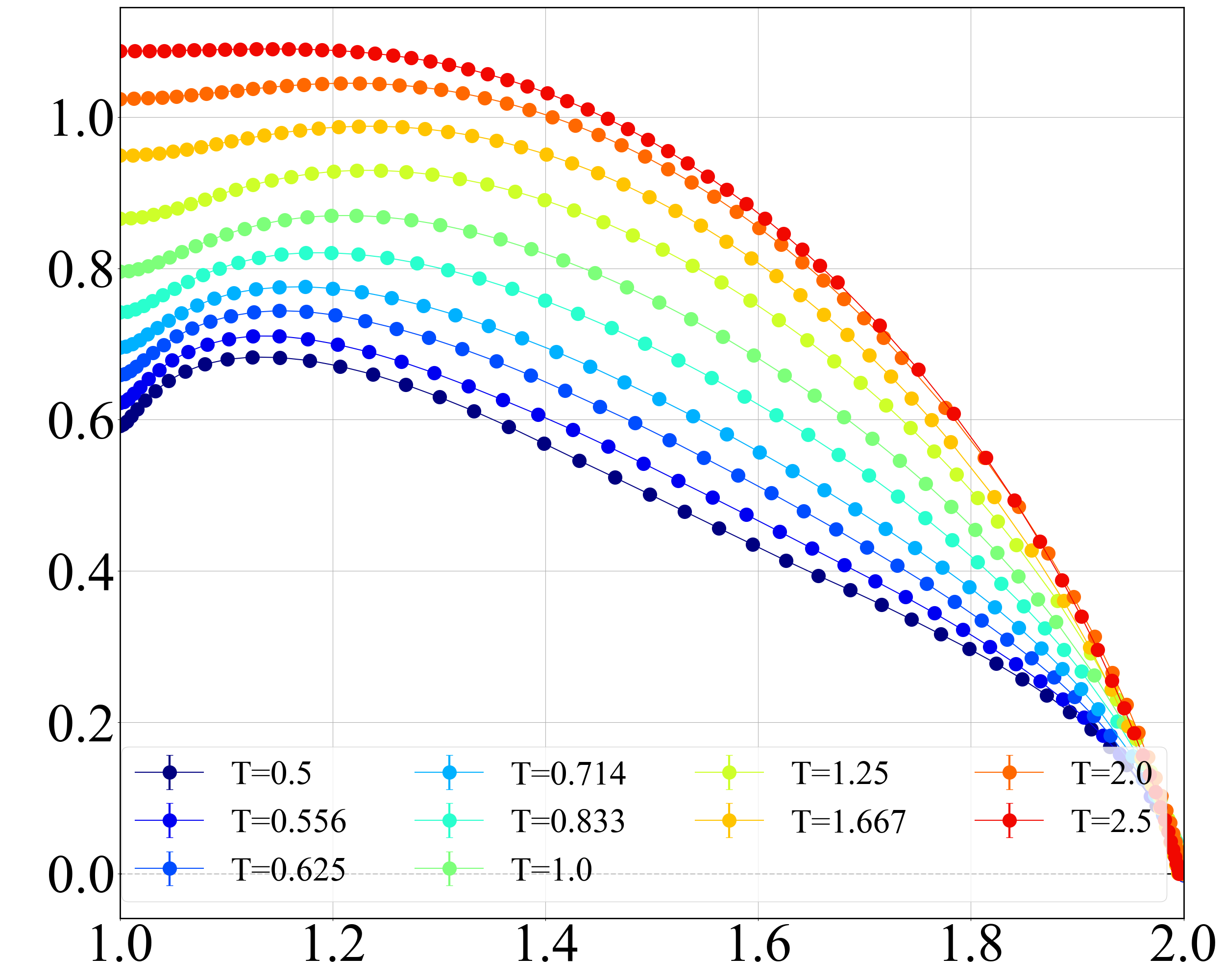}};
\node[left=of img1,node distance=0cm,rotate=90,anchor=center,yshift=-1.0cm,xshift=0.0cm]{\normalsize{ $s[k_B]$}};
\node[below=of img1,node distance=0cm,yshift=1.2cm,xshift=0.0cm]{\normalsize{$n$}};
\node[left=of img1,node distance=0cm,yshift=1.4cm,xshift=5.6cm]{\small{(a)}};
\node[left=of img1,node distance=0cm,yshift=-0.6cm,xshift=4.0cm]{\scriptsize{U=8.0}};
\node (img2) [right=of img1,xshift=-0.75cm] {\includegraphics[width=4.8cm,height=3.8cm]{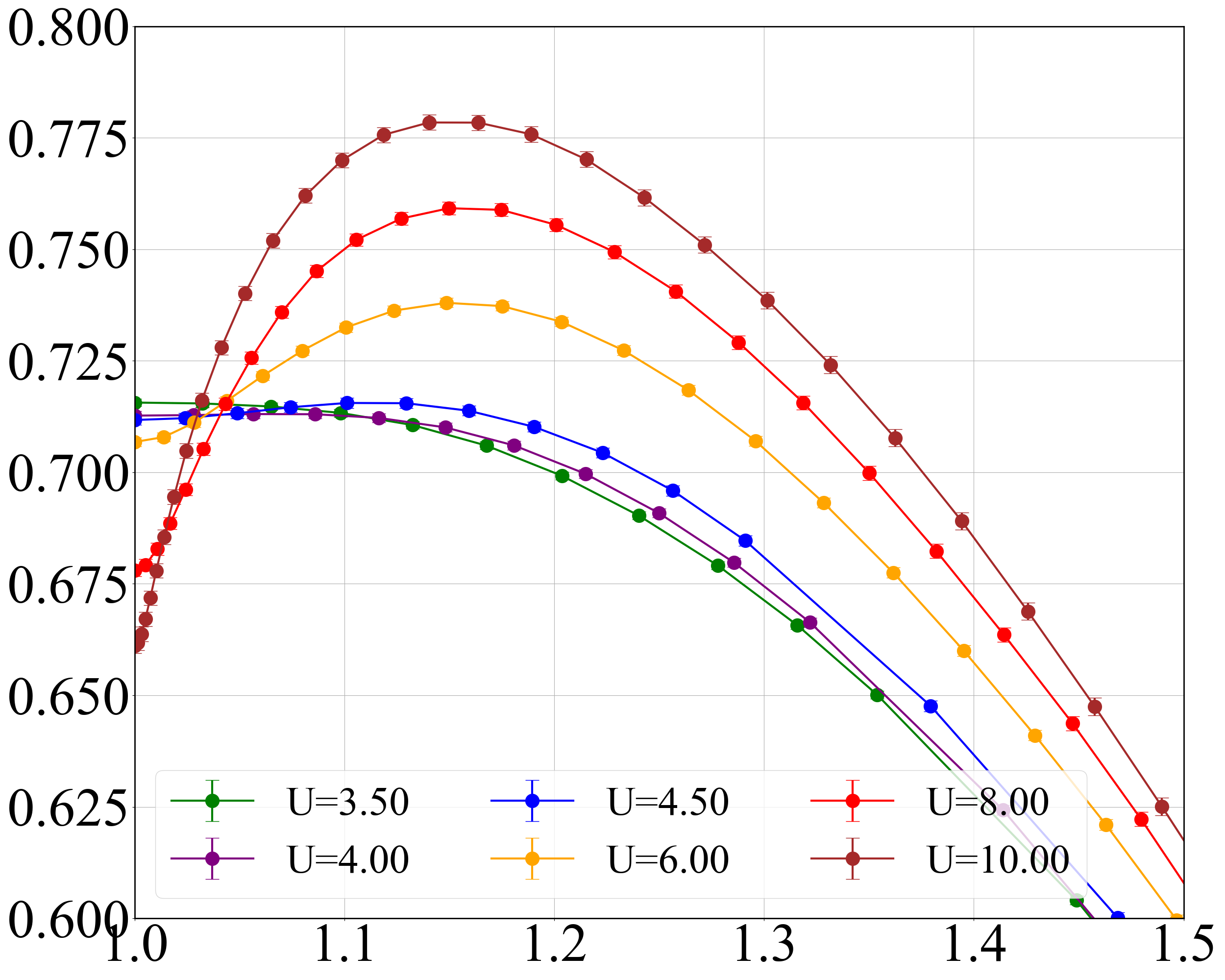}};
\node[left=of img2,node distance=0cm,rotate=90,anchor=center,xshift=0.0cm,yshift=-0.9cm]{\normalsize{ $s[k_B]$}};
\node[below=of img2,node distance=0cm,yshift=1.2cm,xshift=0.0cm]{\normalsize{$n$}};
\node[left=of img2,node distance=0cm,yshift=1.4cm,xshift=5.6cm]{\small{(b)}};
\node[left=of img2,node distance=0cm,yshift=-0.6cm,xshift=4.0cm]{\scriptsize{T=0.67}};
\node (img3) [right=of img2,xshift=-0.75cm]{\includegraphics[width=4.7cm,height=3.8cm]{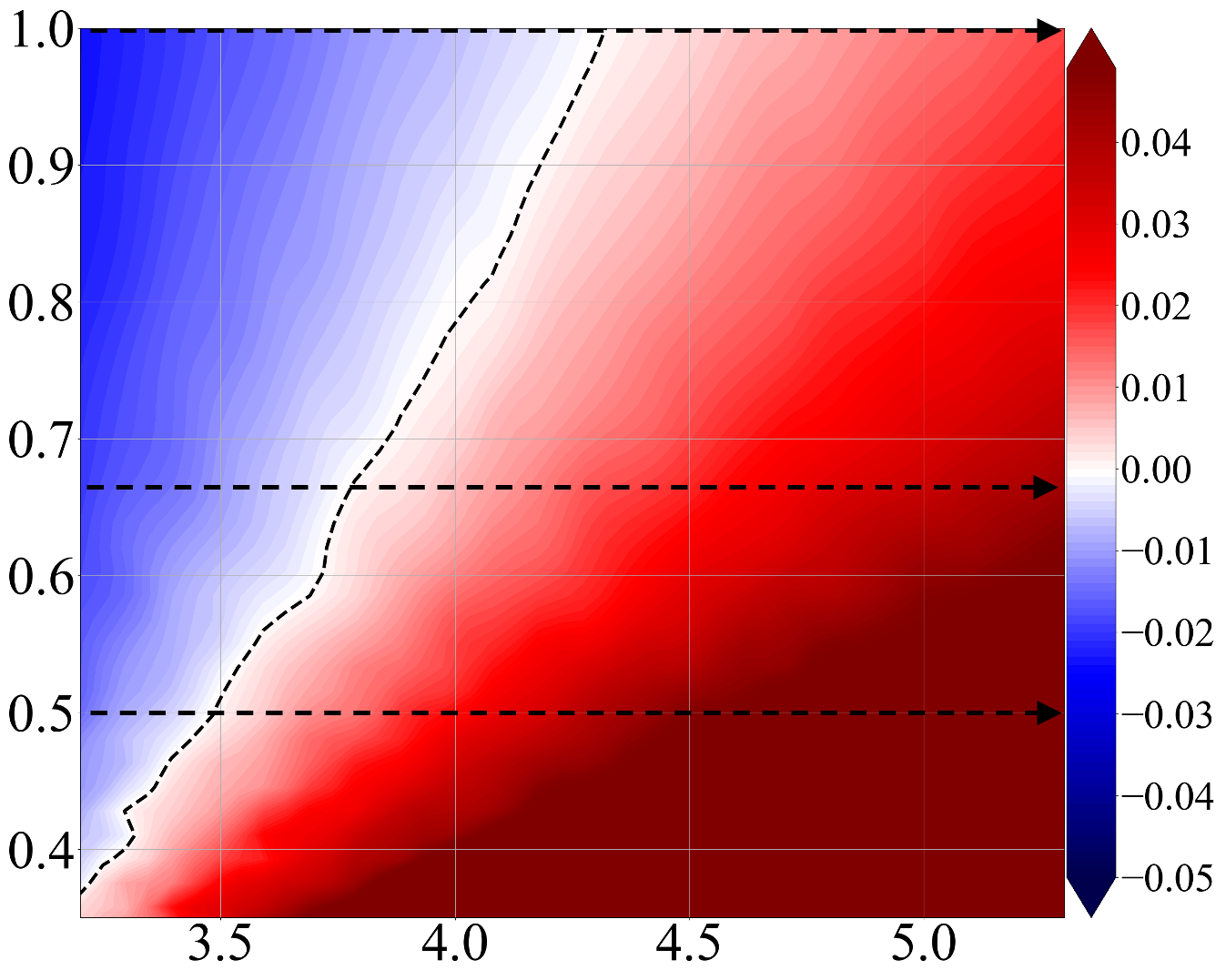}};
\node[above=of img3,node distance=0cm,yshift=-1.2cm,xshift=1.9cm]{\scriptsize{$\frac{\partial \tilde{\kappa}}{\partial T}$}};
\node[left=of img3,node distance=0cm,rotate=0,anchor=center,xshift=0.9cm,yshift=0.0cm]{\normalsize{ $T$}};
\node[left=of img3,node distance=0cm,rotate=0,anchor=center,xshift=2.3cm,yshift=1.6cm]{\tiny{ \color{black}{TDOS-gapless}}};
\node[left=of img3,node distance=0cm,rotate=0,anchor=center,xshift=2.2cm,yshift=1.2cm]{\scriptsize{ \color{black}{No Seebeck}}};
\node[left=of img3,node distance=0cm,rotate=0,anchor=center,xshift=1.9cm,yshift=0.2cm]{\small{ \color{black}{M}}};
\node[left=of img3,node distance=0cm,rotate=0,anchor=center,xshift=3.9cm,yshift=-0.4cm]{\small{ \color{black}{I}}};
\node[left=of img3,node distance=0cm,rotate=0,anchor=center,xshift=2.1cm,yshift=1.0cm]{\scriptsize{ \color{black}{anomaly}}};
\node[left=of img3,node distance=0cm,rotate=0,anchor=center,xshift=3.9cm,yshift=0.50cm]{\tiny{ \color{white}{TDOS-gapped}}};
\node[left=of img3,node distance=0cm,rotate=0,anchor=center,xshift=3.9cm,yshift=0.2cm]{\scriptsize{ \color{white}{Seebeck anomaly}}};
\node[below=of img3,node distance=0cm,yshift=1.2cm,xshift=0.0cm]{\normalsize{$U$}};
\node[left=of img3,node distance=0cm,yshift=1.4cm,xshift=5.0cm]{\small{\color{white}{(c)}}};
\node (img4) [below=of img1,yshift=0.75cm] {\includegraphics[width=4.5cm,height=3.5cm]{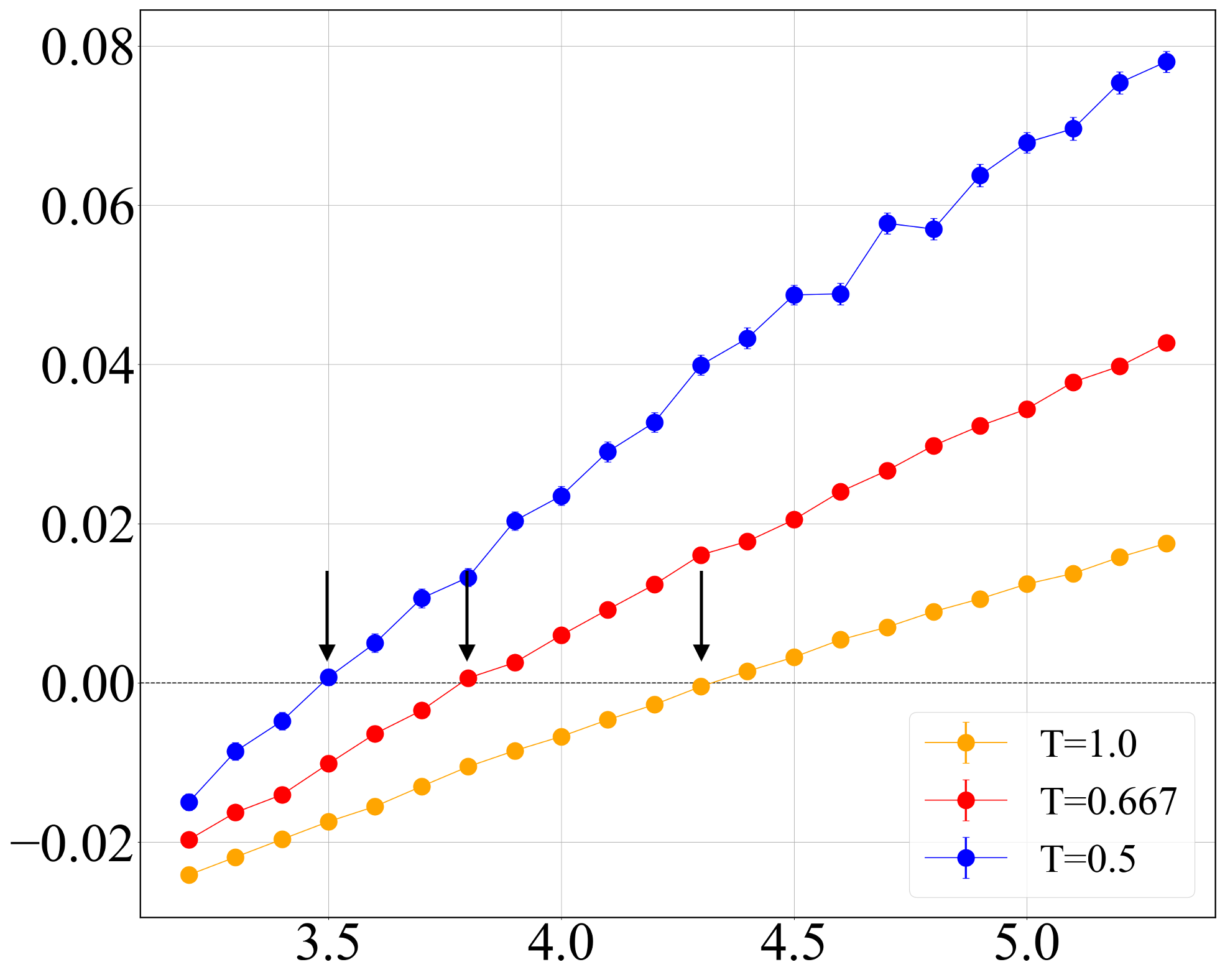}};
\node[left=of img4,node distance=0cm,rotate=0,anchor=center,xshift=0.9cm,yshift=0.0cm]{\large{ $\frac{\partial \tilde{\kappa}}{\partial T}$}};
\node[below=of img4,node distance=0cm,yshift=1.2cm,xshift=0.0cm]{\normalsize{$U$}};
\node[left=of img4,node distance=0cm,yshift=1.4cm,xshift=2.5cm]{\normalsize{(d)}};
\node (img5) [below=of img2,yshift=0.75cm] {\includegraphics[width=4.5cm,height=3.5cm]{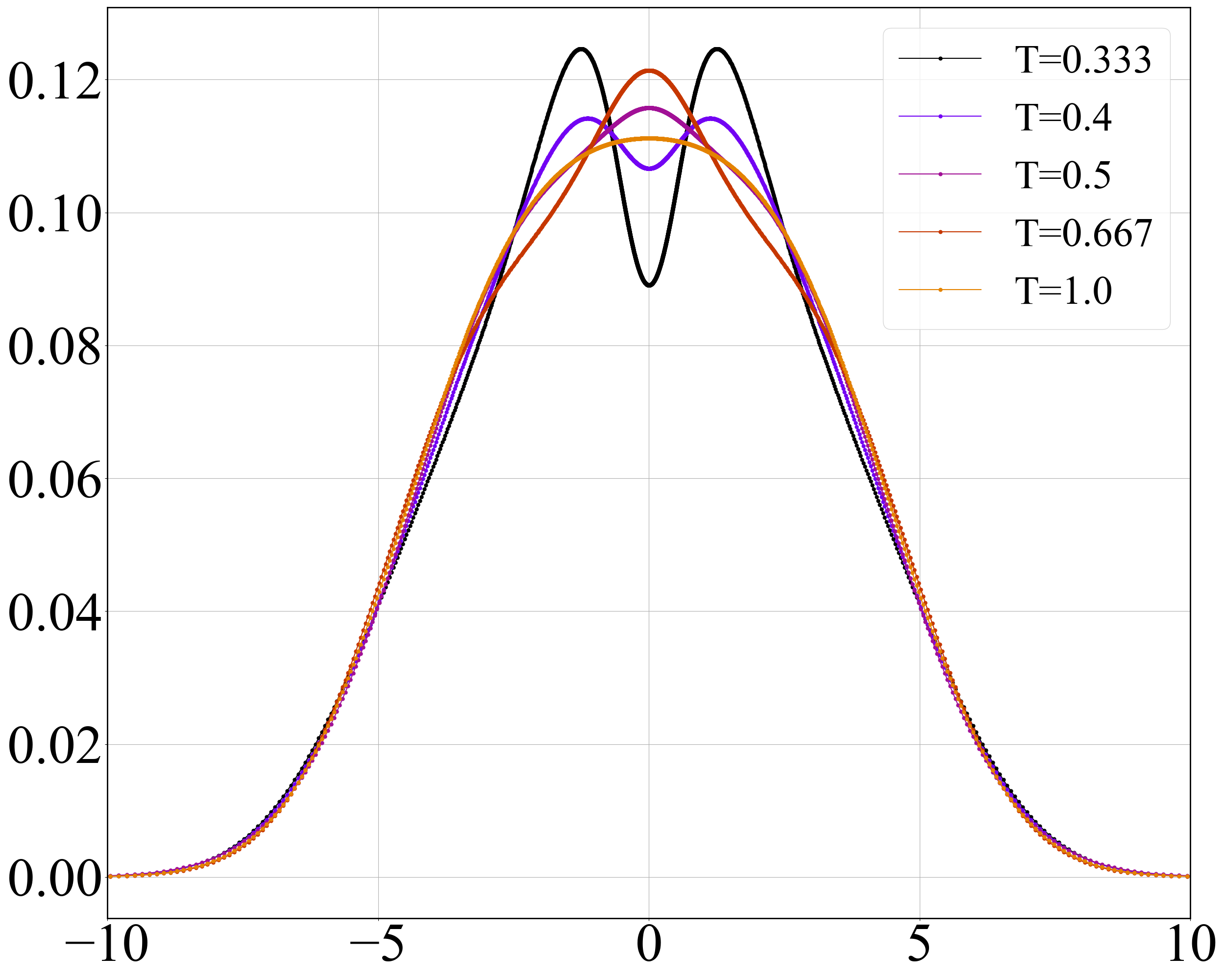}};
\node[left=of img5,node distance=0cm,rotate=90,anchor=center,xshift=0.0cm,yshift=-0.9cm]{\normalsize{ $N(\omega)$}};
\node[below=of img5,node distance=0cm,yshift=1.2cm,xshift=0.0cm]{\normalsize{$\omega$}};
\node[below=of img5,node distance=0cm,yshift=2.2cm,xshift=0.0cm]{\small{$U=4.8$}};
\node[left=of img5,node distance=0cm,yshift=1.4cm,xshift=2.5cm]{\normalsize{(e)}};
\node (img6) [right=of img5,xshift=-0.65cm] {\includegraphics[width=4.2cm,height=3.5cm]{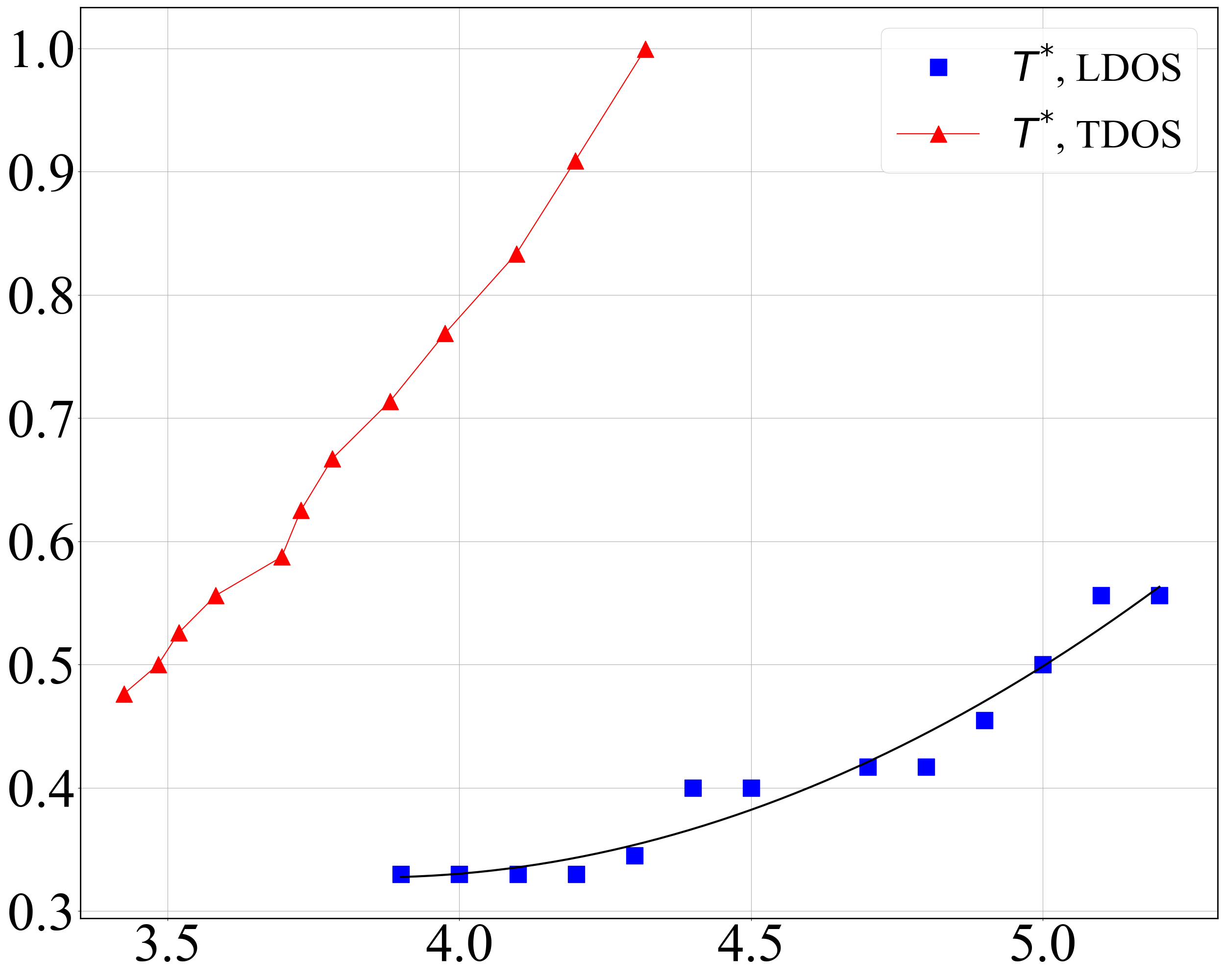}};
\node[left=of img6,node distance=0cm,rotate=0,anchor=center,xshift=0.8cm,yshift=0.0cm]{\normalsize{ $T^{*}$}};
\node[below=of img6,node distance=0cm,yshift=1.2cm,xshift=0.0cm]{\normalsize{$U$}};
\node[left=of img6,node distance=0cm,yshift=1.4cm,xshift=2.5cm]{\small{(f)}};
\end{tikzpicture}
\captionof{figure}{Origin of the Seebeck anomaly: \textbf{(a)} Thermodynamic entropy density at $U=8.0$ with increasing temperature: At low temperatures, there is a peak in the entropy as a function of density close to the Mott Insulating limit. This peak moves towards higher doping with increasing temperature. Beyond a crossover temperature $T_c(U)$, the entropy flattens out with a local maximum at half filling; this temperature coincides with the vanishing of the thermodynamic density of states (TDOS). \textbf{(b)} Entropy as a function of density at $T=0.67$: With increasing $U$, a gap opens in TDOS at $U_{c}(T)$ which pushes the entropy maxima to finite doping. Beyond $U = 6$, the location of the entropy maxima is not sensitive to the strength of $U$ though its value continues to increase. \textbf{(c)} TDOS map {\em at half filling} in the $U,T$ demarcating regions $\frac{\partial \tilde{\kappa}}{\partial T}<0 $ (metallic) from $\frac{\partial \tilde{\kappa}}{\partial T} >0$ (insulating) with the black line showing the separatrix at $\frac{\partial \tilde{\kappa}}{\partial T} = 0$. This provides the evidence that the temperature variation of TDOS {\em at half filling} determines the presence of a maximum in the entropy at a finite doping and hence the sign change of the Seebeck coefficient.
The dotted black lines mark the temperatures for which TDOS($U$) are shown in panel (d).
This maps out a phase diagram for the Seebeck anomaly in terms of TDOS. \textbf{(d)} Zero crossings of $\frac{\partial \tilde{\kappa}}{\partial T}$ defines $U_c(T)$ at a fixed temperature, calculated along dashed arrows in panel (c). The approximate $U_c(T)$ are marked by vertical black arrows. \textbf{(e)} Temperature evolution of the single particle density of states (LDOS), $N(\omega) = -\frac{1}{\pi}{\rm Im} G(r=0,\omega)$, evaluated at half filling. Below a certain temperature, a gap forms at $\omega = 0$. \textbf{(f)} Temperature $T^{*}$ at which a gap opens in TDOS contrasted with temperature at which a pseudogap develops in LDOS. For $U\sim[3.2,3.8]$, no pseudogap formation was found up to the lowest temperatures considered here ($T=0.33$).}
\label{Entropy_charge_gap}
\end{figure*}

\textit{Onset of Seebeck anomaly:} 
The Kelvin formula allows us to relate the Seebeck coefficient to the thermodynamic entropy (defined per unit area), $s = \frac{1}{T}(\epsilon_k+\epsilon_p-\mu n)$, where $\epsilon_k, \epsilon_p$ are the kinetic and potential energy densities respectively, and $n$ is the number density. In the non-interacting limit, the entropy is maximum at half filling. For $U<U_c(T)$ the entropy retains its maximum value at half filling. However, for stronger interactions $U>U_c(T)$, the maximum in the entropy shifts to a finite doping, see Fig.~\ref{Entropy_charge_gap}(b); furthermore the peak value continues to shift to higher dopings with increasing temperature initially,  shown in Fig.~\ref{Entropy_charge_gap}(a). 

It is evident from the Maxwell relation, $\frac{\partial ^2 s}{\partial \mu^2} = \frac{\partial \tilde{\kappa}}{\partial T}$ that a maximum of the entropy at finite doping requires $\frac{\partial \tilde{\kappa}}{\partial T} = 0$ at some chemical potential $\mu$, and increasing the interaction strength $U$ shifts the entropy away from half filling~\cite{lenihan2021entropy}. 
At the critical strength $U_c(T)$, $\frac{\partial \tilde{\kappa}}{\partial T}=0$ appears at the half filling point, signifying a crossover from a metal below $U_c(T)$ to an insulator above $U_c(T)$~\cite{roy2024signatures,kim2020spin}. 

The maximum in entropy as a function of $n$ identifies the location of the sign changes $n_s(T,U)$ of the Seebeck coefficient $S_{\rm Kelvin}$ for a given $T$ and $U$. We argue below that we can determine if a particular set of parameters $T,U$ will have a sign change only at the expected half filling or will also have the anomalous sign change, by considering the behavior of the thermodynamic density of states (TDOS) $\tilde\kappa=\frac{dn}{d\mu}$ {\em at half filling}. 

Fig.~\ref{Entropy_charge_gap}(c) shows a phase diagram in the $T-U$ plane separating two regions: Region M with $\frac{\partial \tilde{\kappa}}{\partial T} <0$, from Region I with $\frac{\partial \tilde{\kappa}}{\partial T} >0$, with a separatrix marking the boundary $\frac{\partial \tilde{\kappa}}{\partial T} =0$.  Region M is a metallic state where there is no gap in the TDOS at the chosen $T,U$ and Region I is an insulating state with a pseudogap or a thermally activated density fluctuations in the TDOS at the chosen $T,U$. This definition above allows us to extend the concepts of metal and Mott insulator to finite temperatures. 
Importantly, these regions also coincide with the absence of a Seebeck anomaly in the metallic regions of the phase diagram and conversely, the presence of a Seebeck anomaly in the insulating regions, due to the Maxwell construction stated above. In Appendix~\ref{Seebeck_MIT}, we explicitly show this case by following the evolution of $S_{\rm kelvin}$ with increasing $U$.

\begin{figure}
\begin{tikzpicture}
\node (img1) {\includegraphics[width=3.7cm,height=3.0cm]{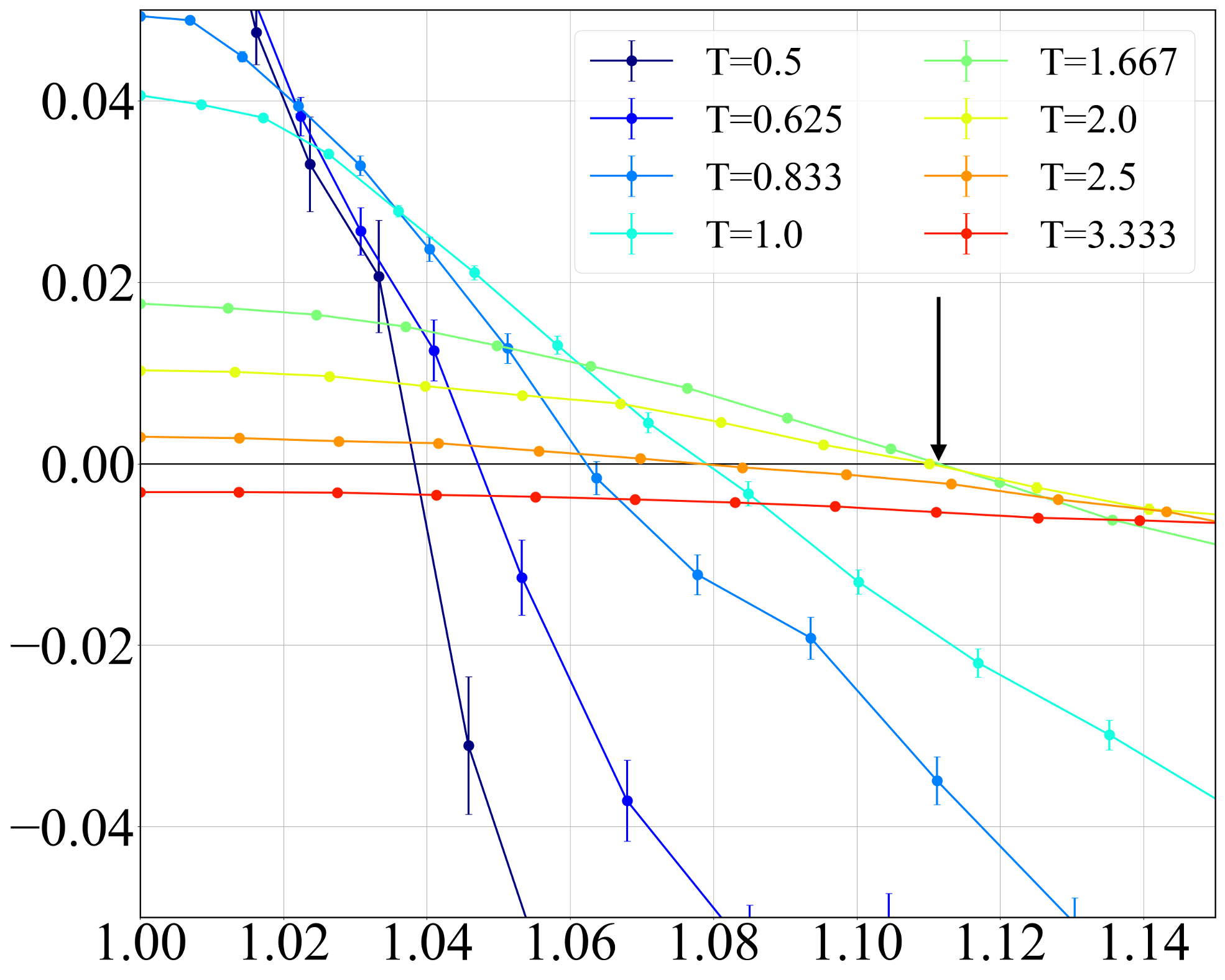}};
\node[left=of img1,node distance=0cm,rotate=0,anchor=center,xshift=0.9cm,yshift=0.0cm]{\normalsize{ $\frac{\partial \tilde{\kappa}}{\partial T}$}};
\node[below=of img1,node distance=0cm,yshift=1.2cm,xshift=0.0cm]{\normalsize{$n$}};
\node[left=of img1,node distance=0cm,yshift=0.3cm,xshift=4.6cm]{\tiny{$n_c$}};
\node[left=of img1,node distance=0cm,yshift=-1.1cm,xshift=2.4cm]{\small{(a)}};
\node (img2) [right=of img1,xshift=-0.65cm]{\includegraphics[width=3.7cm,height=3.0cm]{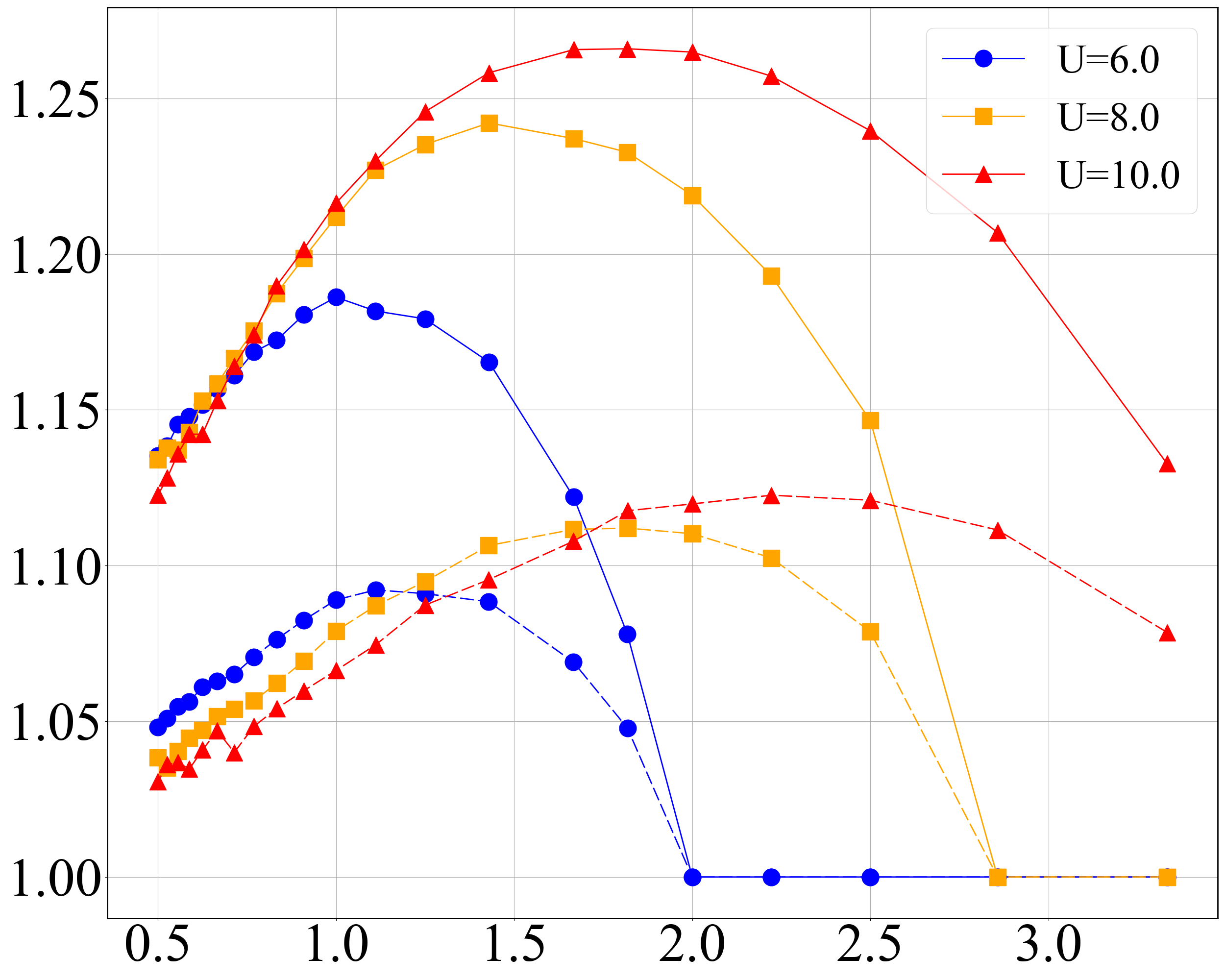}};
\node[left=of img2,node distance=0cm,rotate=90,anchor=center,yshift=-0.9cm,yshift=0.0cm]{\normalsize{ n$_{c,s}$}};
\node[below=of img2,node distance=0cm,yshift=1.2cm,xshift=0.0cm]{\normalsize{$T$}};
\node[left=of img2,node distance=0cm,yshift=-1.1cm,xshift=2.4cm]{\small{(b)}};
\node (img3) [below=of img1,yshift=0.85cm]{\includegraphics[width=3.7cm,height=3.0cm]{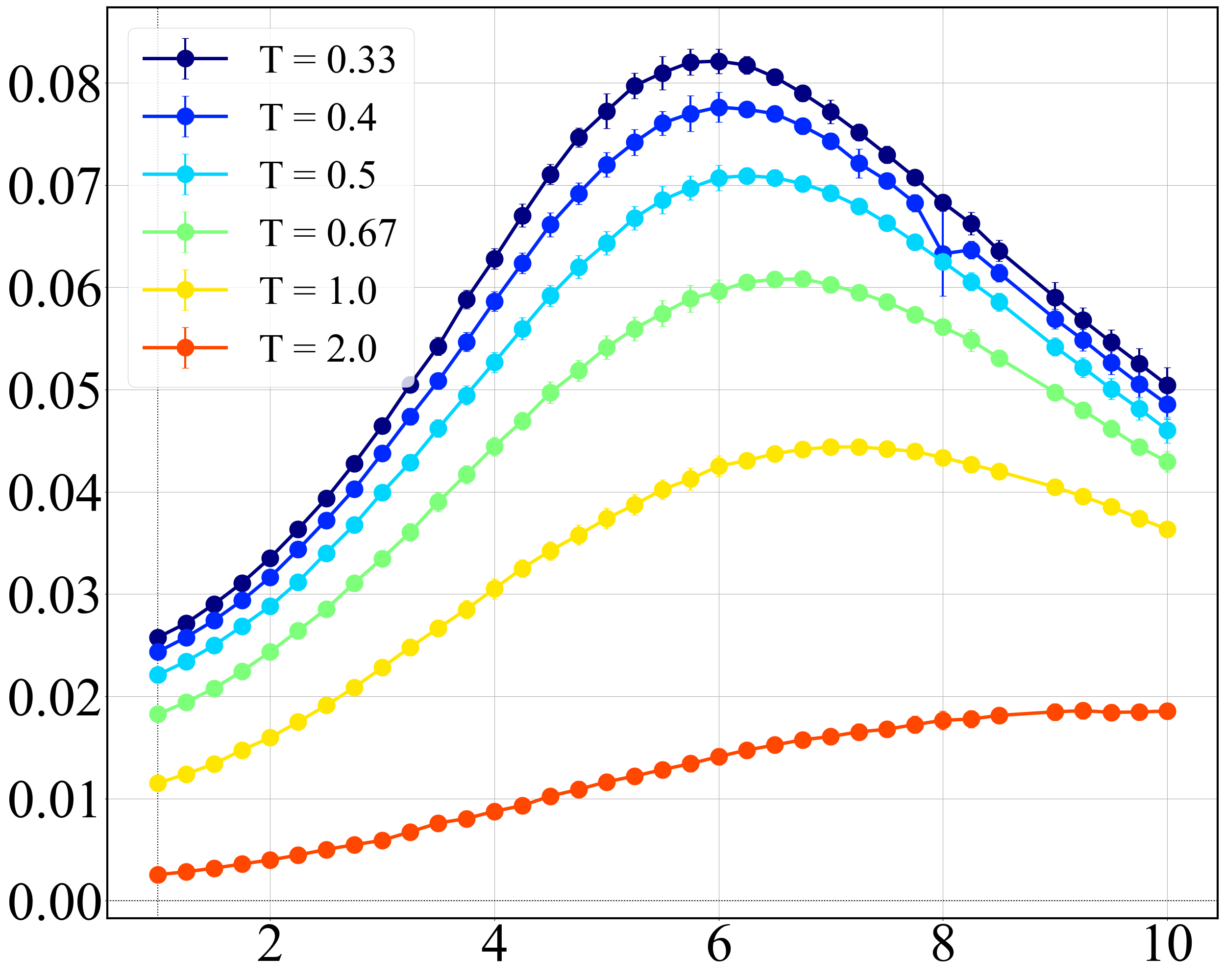}};
\node[left=of img3,node distance=0cm,rotate=90,anchor=center,xshift=0.0cm,yshift=-0.8cm]{\normalsize{ C$_{mm}(nl)$}};
\node[below=of img3,node distance=0cm,yshift=1.2cm,xshift=0.0cm]{\normalsize{$U$}};
\node[left=of img3,node distance=0cm,yshift=-0.4cm,xshift=4.0cm]{\scriptsize{$n=1$}};
\node[left=of img3,node distance=0cm,yshift=-1.0cm,xshift=4.6cm]{\small{(c)}};
\node (img4) [below=of img2,yshift=0.85cm]{\includegraphics[width=3.7cm,height=3.0cm]{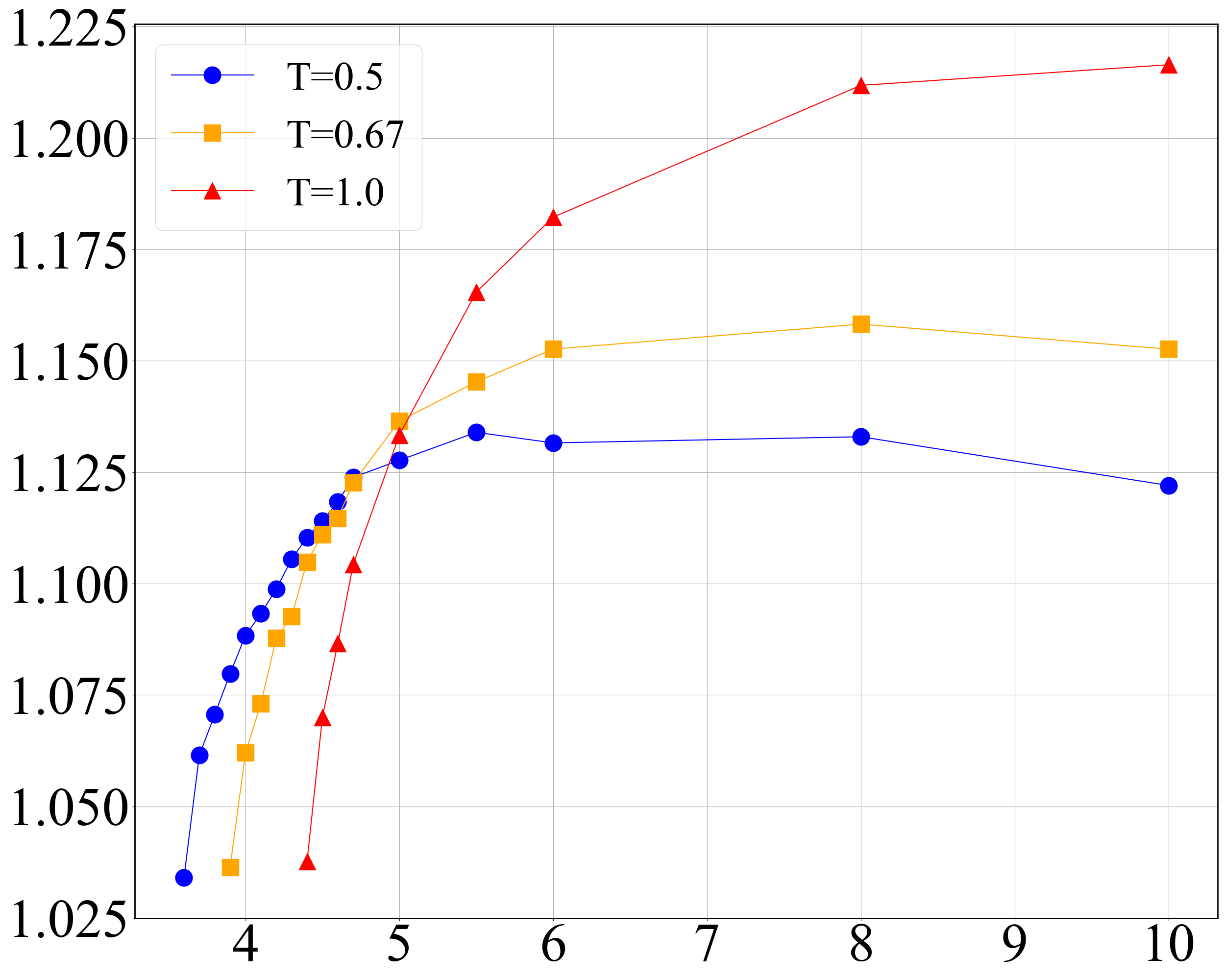}};
\node[left=of img4,node distance=0cm,rotate=0,anchor=center,xshift=0.8cm,yshift=0.0cm]{\normalsize{ n$_{s}$}};
\node[below=of img4,node distance=0cm,yshift=1.2cm,xshift=0.0cm]{\normalsize{$U$}};
\node[left=of img4,node distance=0cm,yshift=-1.1cm,xshift=4.6cm]{\small{(d)}};
\end{tikzpicture}
\captionof{figure}{Approach of the Seebeck coefficient to free particle limit and the insulator-metal crossover. \textbf{(a)} Temperature variation of $\tilde{\kappa}$ tracks the finite doping insulator-metal crossover. The insulating phase (dopings where $\frac{\partial \tilde{\kappa}}{\partial T}>0$) grows with temperature, before turning around and vanishing at a temperature $T_c(U)$ where the charge gap in TDOS closes. \textbf{(b)} Doping level $n_s$ (solid lines) at which S$_{\text{kelvin}}$ changes sign and $n_c$ (dashed lines) at which $\frac{\partial \tilde{\kappa}}{\partial T}$ changes sign (panel a). Both show the same non-monotonic temperature dependence; $n_s$  resets to non-interacting value of $n=1$ at the same temperature where $n_c$ disappears. \textbf{(c)} Non-local contribution to moment moment correlations at half filling. Correlations grow with $U$ in the weak coupling regime until local moments are well formed (peak in $C^{\rm nl}_{mm}$, and decrease beyond that in the strong coupling regime. \textbf{(d)} $n_s$ as a function of $U$ for multiple temperatures. For weak coupling, $n_s(T)$ decreases with temperature; for strong couplings, the temperature dependence is reversed.}
\label{Seebeck_gap_closing}
\end{figure}
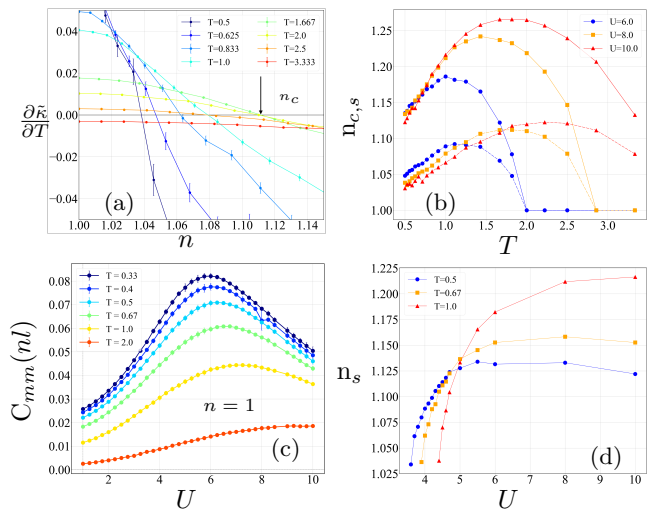

\textit{Insulator-metal crossover, 
Seebeck anomaly and local moments:} From the behavior of TDOS with temperature, one can identify a doping driven Mott Insulator to metal crossover, shown in Fig.~\ref{Seebeck_gap_closing}(a). The Maxwell construction allows one to relate this to the temperature dependence of anomalous zero crossings of $S_{\rm kelvin}$. As shown in Fig.~\ref{Seebeck_gap_closing}(b), the doping at which there is a insulator to metal crossover, $n_c$, closely mirrors the anomalous zero crossings of the Seebeck coefficient $n_s$. It increases initially with temperature upto $T \sim O(t)$, before turning around and disappearing with the Seebeck anomaly when the charge gap in TDOS closes. 



Given that the anomalous Seebeck coefficient is found for strong interaction $U$, we investigate its connection with local moment formation.
The local moment is defined by $m^2_{i} = (n_{i\uparrow}-n_{i\downarrow})^2$ and from that we define the connected moment-moment correlation as $C_{mm}(i,j) = \frac{1}{N_s}\sum_{ij}[\langle m^2_{i}m^2_{j}\rangle - \langle m^2_i\rangle \langle m^2_{j}\rangle]$. $C_{mm}(i,j)$ captures the conditional probability of having a local moment on site $j$, given that site $i$ already has a local moment \cite{cheuk2016observation,hartke2020doublon}. The non-local part of this correlator, $C_{mm}^{\rm nl} = \sum_{i\neq j}C_{mm}(i,j)$ thus serves as a global probe of moment correlations. 

The $U$ dependence of $C_{mm}^{\rm nl}$ at half filling, shown in Fig.~\ref{Seebeck_gap_closing}(c) helps identify two regimes: (i) In the weak coupling regime, moment correlations increase with increasing $U$, peaking at an interaction strength $U_{\rm max}(T)$. The size of the local moments, $m^2$ also grows with $U$ in this regime. (ii) In the strong coupling regime, increasing $U$ beyond $U_{\rm max}(T)$ results in decreasing moment correlations; the size of the local moment starts to saturate on crossover into this regime. Such behavior is also seen at finite doping \cite{roy2024signatures}. $n_s(T,U)$ in Fig.~\ref{Seebeck_gap_closing}(d) differentiates the small $U$ local moment forming regime where $n_s$ is strongly $U$-dependent from the large $U$ regime of well formed local moments where $n_s$ shows minimal $U$ dependence beyond $U_{\rm max}(T)$. The difference between the weak and strong coupling regime, both in terms of $U$ and $T$ dependence can be understood as follows. In the weak coupling regime, increasing temperature washes out the Seebeck anomaly. Since the local moments are not well formed enough, the system tries to minimize free energy $f = \varepsilon_k+\varepsilon_p-Ts$ by lowering the kinetic energy, which competes with moment formation; this effect increases with temperature. In the strong coupling regime, free energy can be minimized by accessing the larger $\ln 2$ entropy from the well formed moments, hence the doping window over which moments can form increases with increasing temperature. At a fixed temperature, once the moments are well formed at $U_{\rm max}(T)$, increasing $U$ contributes weakly to the free energy through $\varepsilon_p = UD$ (where $D$ is the average doublon density). This results in minimal dependence of $n_s(U;T)$ in the strong coupling regime.
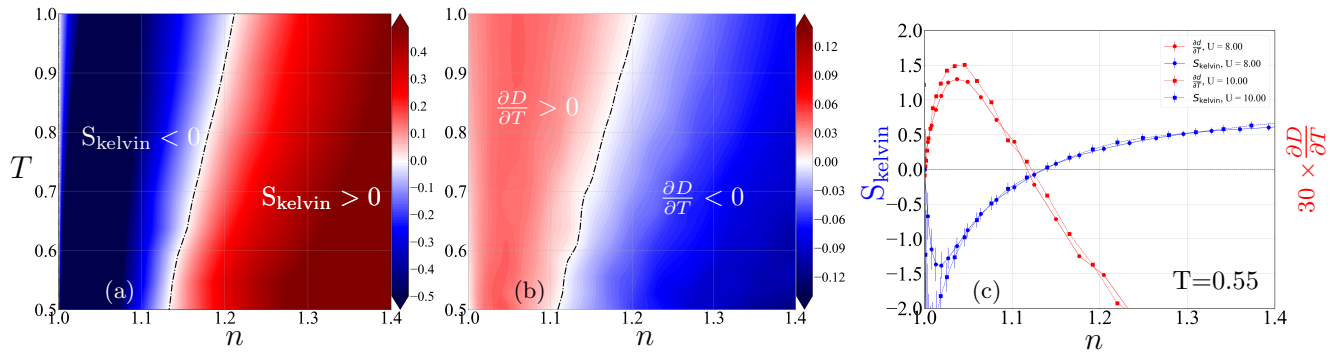
\begin{figure*}
\begin{tikzpicture}
\node (img1) {\includegraphics[width=0.3\linewidth]{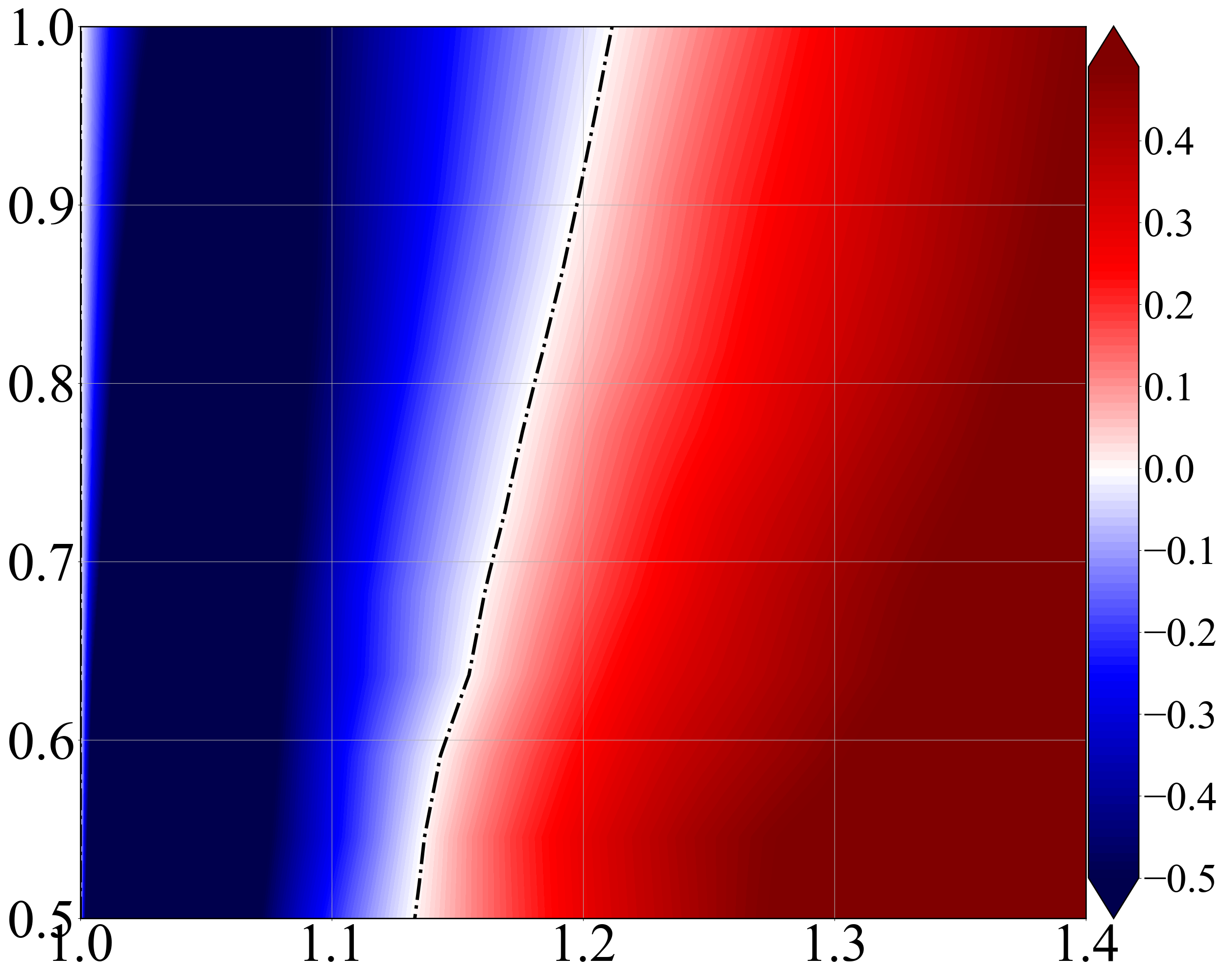}};
\node[left=of img1,node distance=0cm,yshift=-1.65cm,xshift=2.6cm]{\small{\color{white}{(a)}}};
\node[left=of img1,node distance=0cm,rotate=0,anchor=center,xshift=0.9cm,yshift=0.0cm]{\large{ $T$}};
\node[below=of img1,node distance=0cm,yshift=1.2cm,xshift=0.0cm]{\large{$n$}};
\node[left=of img1,node distance=0cm,rotate=0,anchor=center,xshift=4.9cm,yshift=-0.4cm]{\normalsize{ \color{white}{S$_{\text{kelvin}}>0$}}};
\node[left=of img1,node distance=0cm,rotate=0,anchor=center,xshift=2.5cm,yshift=0.4cm]{\normalsize{ \color{white}{S$_{\text{kelvin}}<0$}}};
\node (img2) [right=of img1,xshift=-1.15cm]{\includegraphics[width=0.3\linewidth]{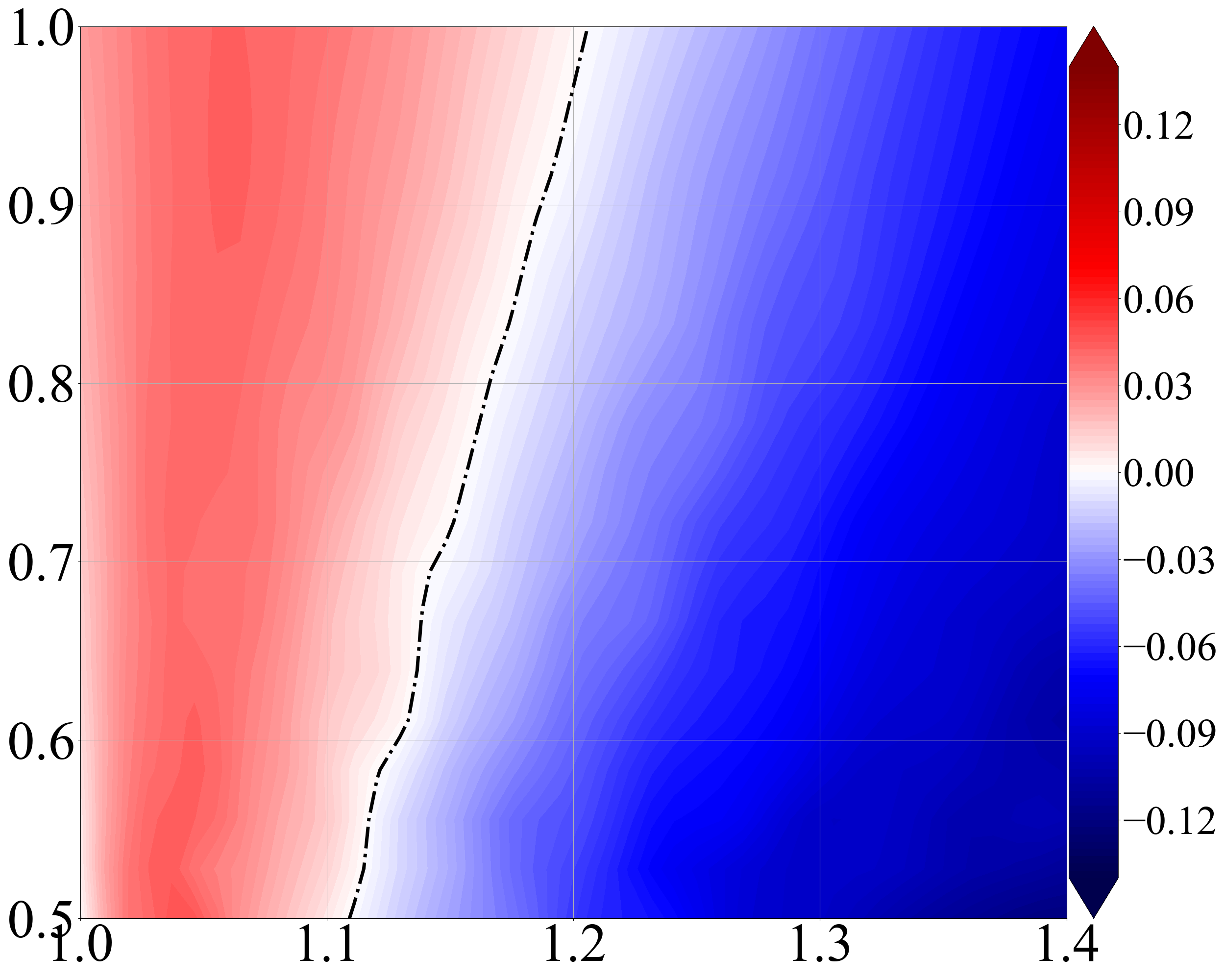}};
\node[left=of img2,node distance=0cm,yshift=-1.65cm,xshift=2.6cm]{\small{(b)}};
\node[below=of img2,node distance=0cm,yshift=1.2cm,xshift=0.0cm]{\large{$n$}};
\node[left=of img1,node distance=0cm,rotate=0,anchor=center,xshift=4.9cm,yshift=-0.4cm]{\normalsize{ \color{white}{S$_{\text{kelvin}}>0$}}};
\node[left=of img2,node distance=0cm,rotate=0,anchor=center,xshift=2.3cm,yshift=0.8cm]{\normalsize{ \color{white}{$\frac{\partial D}{\partial T}>0$}}};
\node[left=of img2,node distance=0cm,rotate=0,anchor=center,xshift=4.5cm,yshift=-0.4cm]{\normalsize{ \color{white}{$\frac{\partial D}{\partial T}<0$}}};
\node (img3) [right=of img2,xshift=-0.95cm]{\includegraphics[width=6cm,height=4.8cm]{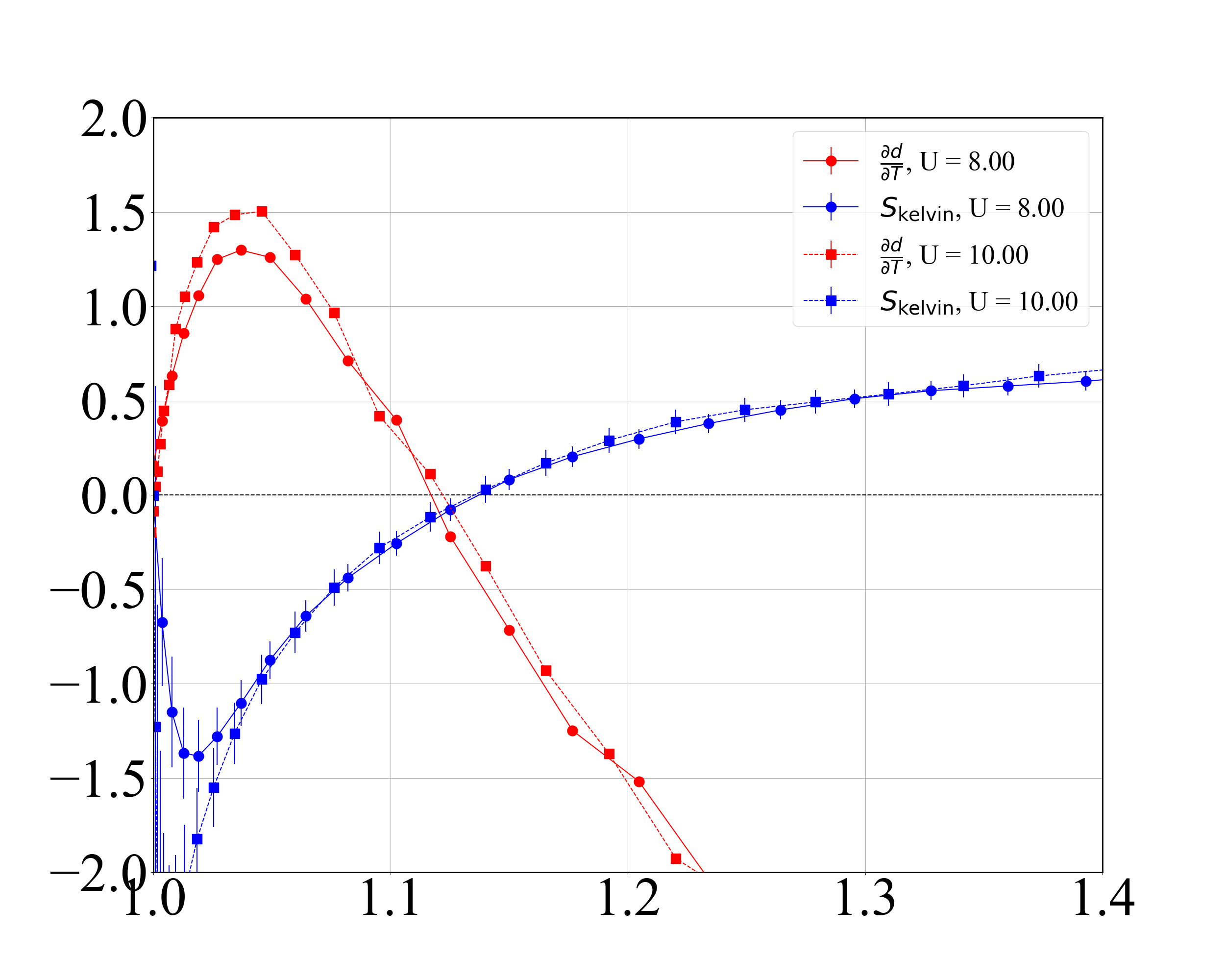}};
\node[left=of img3,node distance=0cm,yshift=-1.65cm,xshift=3.0cm]{\small{(c)}};
\node[left=of img3,node distance=0cm,rotate=90,anchor=center,yshift=-1.2cm,xshift=0.0cm]{\large{ \color{blue}{S$_{\text{kelvin}}$}}};
\node[right=of img3,node distance=0cm,rotate=90,anchor=center,yshift=1.3cm,xshift=0.0cm]{\small{\color{red}{30 $\times$}}\large{\color{red}{$ \frac{\partial D}{\partial T}$}}};
\node[below=of img3,node distance=0cm,yshift=1.4cm,xshift=0.0cm]{\large{$n$}};
\node[right=of img3,node distance=0cm,rotate=0,anchor=center,yshift=-1.5cm,xshift=-2.5cm]{\normalsize{T=0.55}};
\end{tikzpicture}
\captionof{figure}{Seebeck anomaly and local moments. \textbf{(a)} Seebeck coefficient on the $T-n$ plane at $U = 8.0$. The black dashed line is the location of $n_s(T)$, doping at which the Seebeck coefficient changes sign at that temperature. \textbf{(b)} Temperature derivative of the doublon number $D = \langle n_{i\uparrow}n_{i\downarrow} \rangle$. If the doublon number decreases with cooling, the system tends to form local moments and localize. If the doublon number increases with cooling, the system is ``metallic" like with itinerant electrons. \textbf{(c)} Comparison of doping dependence of $\frac{\partial D}{\partial T}$ and Seebeck coefficient for strong interactions at $T = 0.55$. Note that the wrong sign of the Seebeck coefficient is closely associated with the system's tendency to form local moments.}
\label{Seebeck_critical_doping_moments}
\end{figure*}


In Ref. \cite{wang2023quantitative} it is argued that the location of the sign changes of the Seebeck coefficent obtained from the Kubo and Kelvin formulae are similar. 
We therefore propose the saturation of the peaks of entropy, and hence $n_s(U)$, with increasing $U$ at a fixed temperature in the strong coupling regime, as a possible candidate for the universal doping dependence of the Seebeck anomaly seen in experiments~\cite{cooper1987thermoelectric,obertelli1992systematics}.


To further demonstrate that local moments are indeed responsible for the increase of Seebeck anomaly with increasing temperature in the strong coupling regime, we turn to a local probe for moment formation. Fermionic anti-commutation relations enforce that each site can have either a local moment $m^2_i$, a doublon, $d_{i} = \langle n_{i\uparrow}n_{i\downarrow}\rangle$ or a holon, $h_{i} = \langle (1-n_{i\uparrow})(1-n_{i\downarrow}) \rangle$. In the particle doped side, holon occupation is minimal, and doublon and local moment occupation on a single site are hence anti-correlated. 

The temperature variation of the local doublon number has been studied before in context of adiabatic cooling of fermions in optical traps, where a ``Pomeranchuk" like effect can happen depending on the dimensionality of the system \cite{werner2005interaction,paiva2009fermions,dare2007interaction}. Here, we interpret the derivative of the average doublon number $D = (1/N_s)\sum_{i}\langle n_{i\uparrow}n_{i\downarrow}\rangle$, $\frac{\partial D}{\partial T}$ as an indication of localization; decreasing doublon occupation with cooling indicates tendency of the system to pin down electrons to form local moments, while increasing doublon occupation with cooling indicates an itinerant nature of the electrons to form doubly occupied sites. Armed with this, we compare the Seebeck coefficient and $\frac{\partial D}{\partial T}$ in Fig.~\ref{Seebeck_critical_doping_moments}. The zero crossings of the $S_{\rm kelvin}$, in panel (a) is in very close agreement with the zero crossings of $\frac{\partial D}{\partial T}$ in panel (b), where the system goes from having localized electrons to more itinerant electrons. In the strong coupling regime, this is indeed the case irrespective of the interaction strength (Fig.~\ref{Seebeck_critical_doping_moments}(c)). Local moment formation thus dominates the doping window in which the sign of the Seebeck coefficient reverses, and is a key component in driving the anomaly. This was also seen recently in experiments with MATBG~\cite{ghosh2024evidence} due to emergent local moment from flatbands.

\noindent \textit{Conclusion:} We have analyzed the behavior of the anomalous thermopower in the repulsive Hubbard model, a prototype for strongly correlated systems such as cuprates and flatband systems. We have shown that in the incoherent regime, where an entropic representation of the thermopower is valid \cite{mravlje2016thermopower,deng2013bad,arsenault2013entropy}, the anomalous sign change of the Seebeck coefficient is brought on by an opening of the charge gap in the thermodynamic density of states (TDOS). Remarkably, the approach to free particle behavior is highly non-monotonic as a function of temperature; the Seebeck anomaly increases with increasing temperature, before finally turning around at a temperature scale set by the interaction strength, and decreases to the free particle limit at a temperature scale where the charge gap in the TDOS closes. This behavior closely mirrors the doping driven Mott insulator to metallic crossover at the corresponding temperature. We identify the origin of the anomalous phase, where the Seebeck coefficient has a divergence and the ``wrong" sign, is primarily dominated by local moment formation. 

Recently such behavior has also been observed in magic-angle twisted bilayer graphene (MATBLG) \cite{ghosh2024evidence} due to emergent moment formation arising from flatbands, solidifying our interpretation of the role of local moments on anomalous transport behavior in strongly correlated systems. It would be interesting to test our prediction on the non-monotonic dependence of the density on temperature at which the Seebeck coefficient changes sign. Our study provides an understanding of behavior of Seebeck coefficient as seen in experiments on cuprates, and highlights the role of thermodynamic many-body quantities in capturing transport in the incoherent regime. It should be noted that since our results are based on thermodynamic arguments, they are quite general and should hold independently in any system with strong correlations exhibiting metal-insulator crossover and possessing particle-hole symmetry. 

\textit{Acknowledgments:} S.R., A.S. and N.T. acknowledge support from NSF Materials Research Science and Engineering Center (MRSEC) Grant No. DMR-2011876 and NSF-DMR 2138905. Computations were performed at the Unity cluster of Arts and Science College, Ohio State University.

\appendix

\section{Parton mean field theory}
\label{Parton MFT}
In this section, we provide a parton mean field description for the calculation of the Seebeck coefficient. We use a canonical transformation, known as the Schrieffer-Wolff transformation, and derive a low energy effective Hamiltonian for the Hubbard model in the limit $U/t\gg 1$ by eliminating high energy processes order by order~\cite{macdonald1988t}. We start with the repulsive Hubbard model
\begin{align}
H &= -t\sum_{\langle ij \rangle,\sigma}
\lt(\hat{c}^{\dagger}_{i\sigma}\hat{c}_{j\sigma}+\text{H.c}\rt) 
+ U\sum_{i}\hat{n}_{i\up}\hat{n}_{i\dn} -\mu\sum_i \hat{n}_i \nonumber \\
&\equiv H_T + H_0 \; ,
\label{eq:Hub}
\end{align}
where $H_0$ corresponds to the local interaction term which keeps the states within the same energy sector, and $H_T$ is the hopping term which can connect the states between different energy sectors. $H_T$ is separated into three pieces,
\begin{align}
H_T &= T_0+T_1+T_{-1} \; .
\end{align}
$T_n(ij)$ hops a fermion from site $j$ to site $i$, where the total number of double occupancy due to this process increases by $n$, i.e.
\begin{align}
[H_0,T_n] &= nUT_n \; .
\end{align} 
The effective low-energy Hamiltonian is formally obtained by eliminating terms coupling between the low-energy subspace and the high-energy subspace up to second order,
\begin{equation}
H_{\rm eff} = H_0 + T_{0}+\frac{1}{U}[T_{1},T_{-1}]+\mathcal{O}\lt(\frac{t^3}{U^2}\rt).
\label{eq:Heff1}
\end{equation}
\begin{figure}
\centering 
\includegraphics[width=0.75\columnwidth]{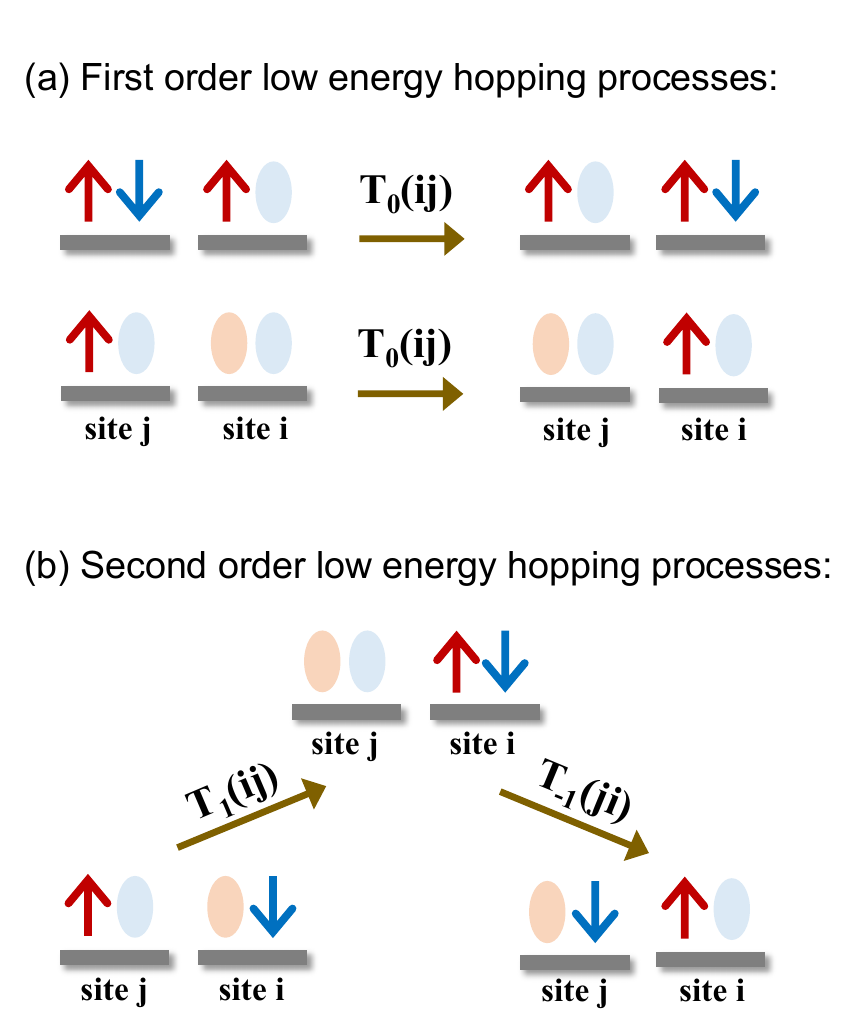}
\caption{ First order and second order low energy hopping processes appearing  in the effective Hamiltonian~\ref{eq:Heff1}. (a) In first order, $T_0$ hops a spinon from one site to its nearest neighbor site, where the doublon number remains the same. Two cases are shown for the hopping of an up spin with the doublon numbers 0 and 1, respectively. (b) In second order, the low energy hopping process $[T_1,T_{-1}]$ involves the flipping of two spins on two nearest neighbor sites, via the formation of a virtual \textit{high energy} double occupancy at the intermediate state. }
\label{fig:MsmallT}
\end{figure}
Next we invoke a parton mean field theory, where the electron operators are decomposed into spinful fermionic operators (spinons) and spinless bosonic operators (doublons and holons)~\cite{kopp1988superconductivity}:
\begin{align}
\hat{c}^{\dagger}_{i\sigma} & = \hat{f}^{\dagger}_{i\sigma}\hat{h}_{i}+\sigma\hat{f}_{i\bar{\sigma}}\hat{d}^{\dagger}_{i} \; . 
\end{align}
The physical Hilbert space is restored by imposing the following constraint on every site:
\begin{align}
\hat{d}^{\dagger}_{i}\hat{d}_{i}+\hat{h}^{\dagger}_{i}\hat{h}_{i}+\sum_{\sigma}\hat{f}^{\dagger}_{i\sigma}\hat{f}_{i\sigma} = 1 \; ,
\label{constraint_1}
\end{align}
which is implemented in the Hamiltonian by introducing a Lagrange multiplier $\lambda$ at the mean-field level (i.e. \textit{on average}). Including all of these, the effective Hamiltonian (Eq.~\ref{eq:Heff1}) using the parton operators can be written as
\begin{align}
&H_{\rm eff} = U\sum_i \hat{d}^\dagger_i\hat{d}_i
-t\sum_{\langle ij\rangle,\sigma}\lt(\hat{d}^\dagger_i \hat{d}_j \hat{f}_{i\sigma} \hat{f}^\dagger_{j\sigma}+\hat{h}_i\hat{h}^\dagger_j \hat{f}^\dagger_{i\sigma} \hat{f}_{j\sigma}\rt) \nonumber \\
\label{eq:Heff}
&+ J\sum_{\langle ij\rangle} \!\lt(2\hat{n}^d_i \hat{n}^h_j - \sum_\sigma \hat{n}_{i\sigma} \hat{n}_{j\bar{\sigma}} + \sum_{\sigma}\hat{f}^\dagger_{i\sigma}\hat{f}_{i\bar{\sigma}}\hat{f}^\dagger_{j\bar{\sigma}}\hat{f}_{j\sigma}\!\rt) \\
&\!\!\!\! -\!\mu\!\sum_i\!\lt(\!2\hat{d}^\dagger_i\hat{d}_i\!+\!\!\sum_\sigma \!\hat{f}^\dagger_{i\sigma}\hat{f}_{i\sigma}\!\!\rt) \!-\! \lambda\!\sum_{i}\!\lt(\!\hat{d}^\dagger_i\hat{d}_i\!+\!\hat{h}^\dagger_i\hat{h}_i\!+\!\!\sum_\sigma \hat{f}^\dagger_{i\sigma}\hat{f}_{i\sigma}\!\!\rt) \nonumber
\end{align}
where $J=4t^2/U$.
Next, we consider the following mean field order parameters:
\begin{align}
n_{d} &= \frac{1}{N_s}\sum_{i}\langle \hat{d}^\dagger_i\hat{d}_i\rangle, ~~ n_{h}=\frac{1}{N_s}\sum_i\langle\hat{h}^{\dagger}_{i}\hat{h}_{i}\rangle \nonumber \\
n_{f} &=\frac{1}{2N_s}\sum_{i,\sigma}\langle \hat{f}^{\dagger}_{i\sigma}\hat{f}_{i\sigma} \rangle, ~~~ \chi_{d} =\frac{1}{zN_s}\sum_{\langle ij\rangle}\langle \hat{d}^{\dagger}_{i}\hat{d}_{j} \rangle \nonumber \\
\chi_{h} &=\frac{1}{zN_s} \sum_{\langle ij\rangle}\langle \hat{h}^{\dagger}_{i}\hat{h}_{j} \rangle, ~~~ \chi_{f} = \frac{1}{2zN_s}\sum_{\langle ij\rangle,\sigma}\langle\hat{f}^{\dagger}_{i\sigma}\hat{f}_{j\sigma}\rangle
\label{Mean_field_order_parameter}
\end{align}
and obtain the mean field Hamiltonian
\begin{align}
H^{\rm MF}_{\rm eff} \!=\! \sum_k \!\lt(\!E_d(k)\hat{d}^\dagger_k\hat{d}_k + E_h(k)\hat{h}^\dagger_k\hat{h}_k + \sum_\sigma E_f(k)\hat{f}^\dagger_{k\sigma}\hat{f}_{k\sigma}\!\rt)
\end{align}
with the bosonic and the fermionic energies given by
\begin{align}
E_d(k) &= U-2\mu-\lambda+2Jzn_h+2t\gamma(k)\chi_f \nonumber \\
E_h(k) &= -\lambda+2Jzn_d-2t\gamma(k)\chi_f \\
E_f(k) &= -\mu -\lambda+t\gamma(k)(\chi_d-\chi_h)-2J\gamma(k)\chi_f-2Jzn_f \nonumber
\label{Mean_field_hamiltonian}
\end{align}
with $\gamma(k) = 2[\cos(k_x)+\cos(k_y)]$ and $z=4$ is the coordination number in two dimensions. We first solve the mean field equations self-consistently on a square lattice and calculate the Seebeck coefficient using the Kelvin formula, as described in the Main text.
\begin{figure}
\begin{tikzpicture}
\node (img1) {\includegraphics[width=0.55\linewidth]{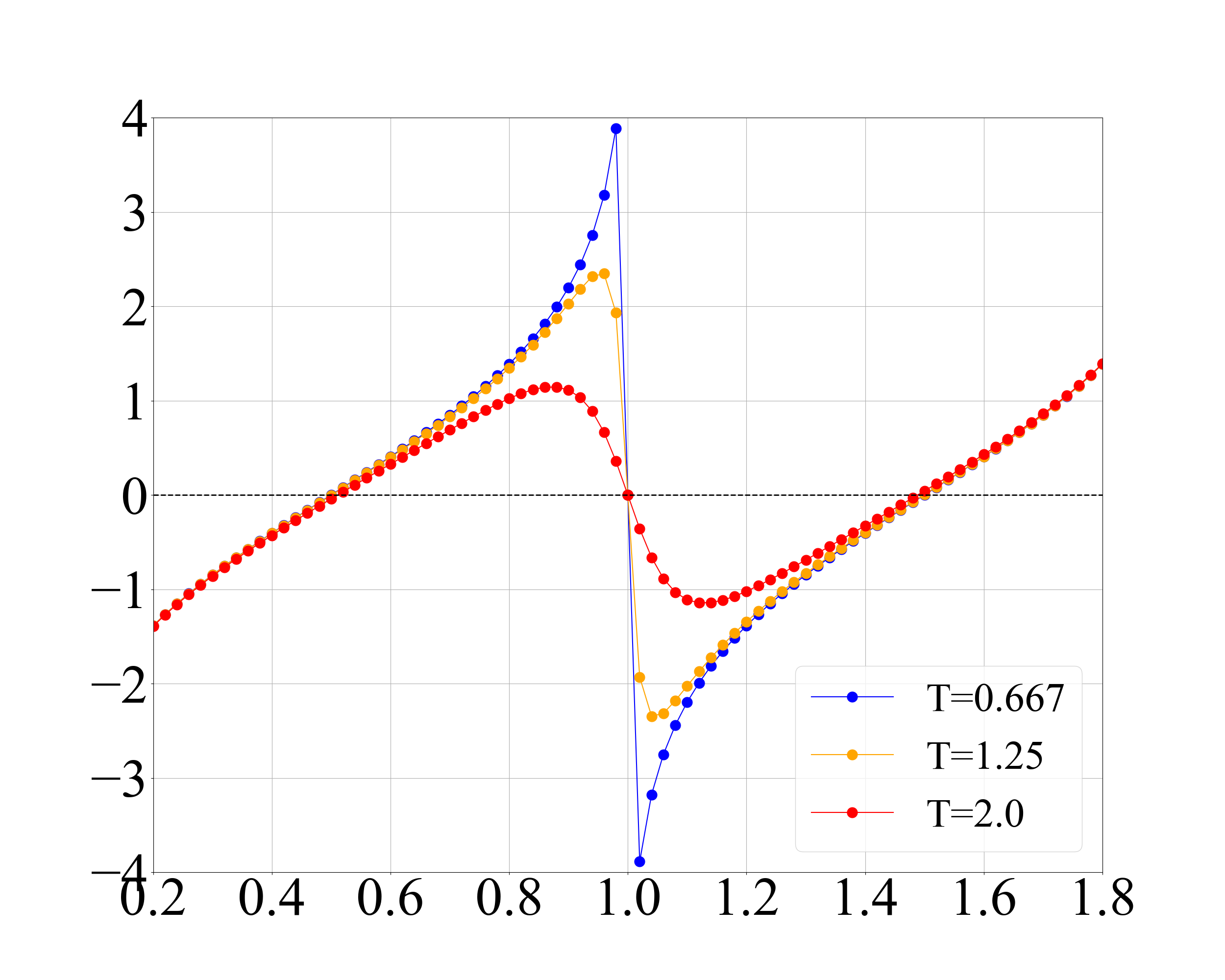}};
\node[left=of img1,node distance=0cm,rotate=90,anchor=center,yshift=-1.3cm,xshift=0.0cm]{\small{ S$_{\text{kelvin}}[k_B/e]$}};
\node[left=of img2,node distance=0cm,yshift=-1.35cm,xshift=2.6cm]{\small{(a)}};
\node[below=of img1,node distance=0cm,yshift=1.2cm,xshift=0.0cm]{\large{$n$}};
\node[left=of img1,node distance=0cm,yshift=-1.25cm,xshift=2.6cm]{\small{(a)}};
\node[left=of img1,node distance=0cm,yshift=1.2cm,xshift=3.2cm]{\footnotesize{$U=10.0$}};
\node (img2) [right=of img1,xshift=-1.35cm]{\includegraphics[width=0.55\linewidth]{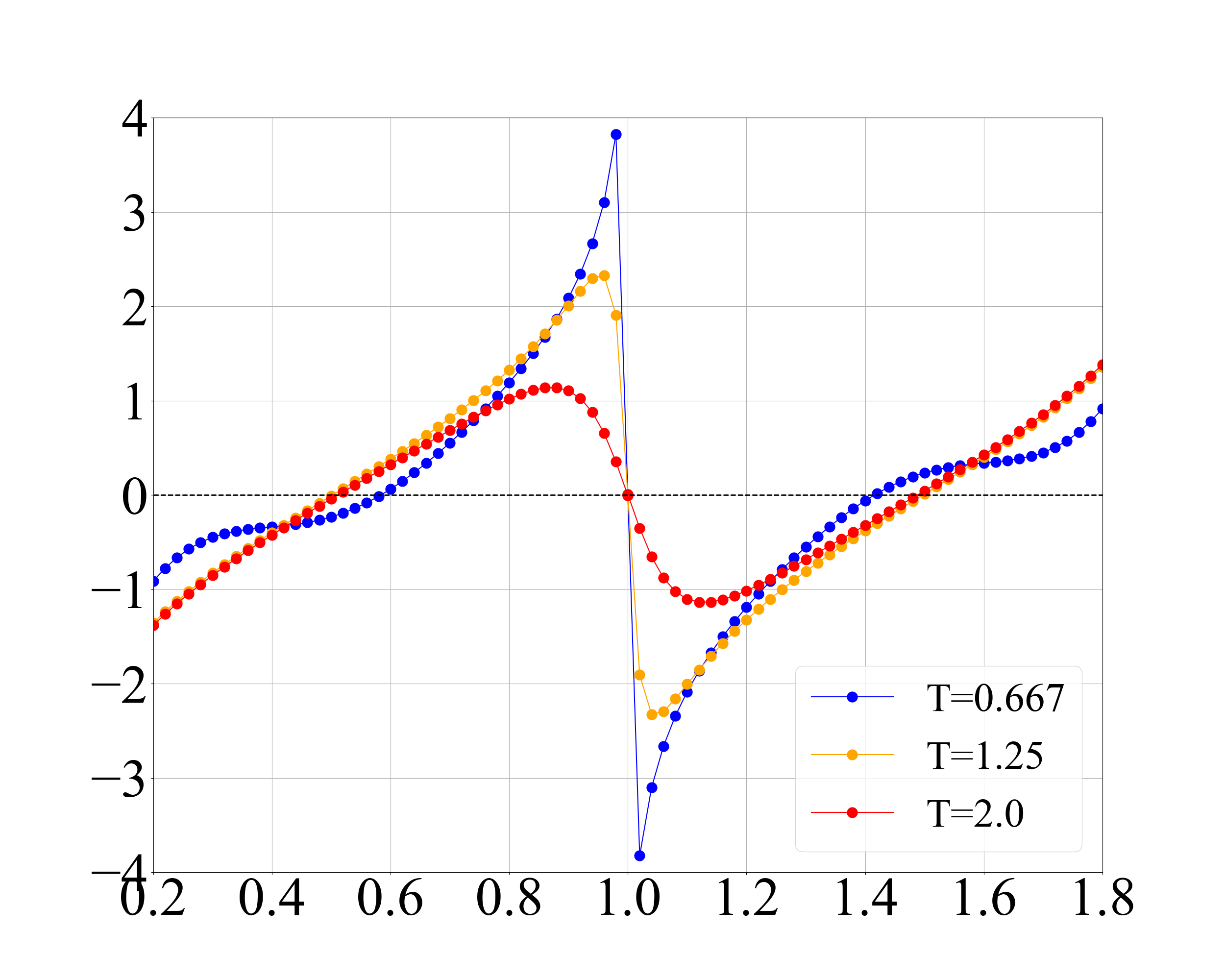}};
\node[left=of img2,node distance=0cm,yshift=-1.25cm,xshift=2.6cm]{\small{(b)}};
\node[left=of img2,node distance=0cm,rotate=90,anchor=center,yshift=-1.3cm,xshift=0.0cm]{\small{ S$_{\text{kelvin}}[k_B/e]$}};
\node[left=of img2,node distance=0cm,yshift=1.2cm,xshift=3.2cm]{\footnotesize{$U=10.0$}};
\node[below=of img2,node distance=0cm,yshift=1.2cm,xshift=0.0cm]{\large{$n$}};
\end{tikzpicture}
\captionof{figure}{Seebeck coefficient calculated from the parton construction of the effective Hamiltonian Eq.~\ref{eq:Heff}, at $U = 10.0$. \textbf{(a)} Seebeck coefficient obtained by keeping only the density terms $n_d$, $n_h$, $n_f$. \textbf{(b)} Seebeck coefficient obtained by keeping density terms and nearest neighbor hopping terms $\chi_d$, $\chi_h$, $\chi_f$.}
\label{Seebeck_parton}
\end{figure}
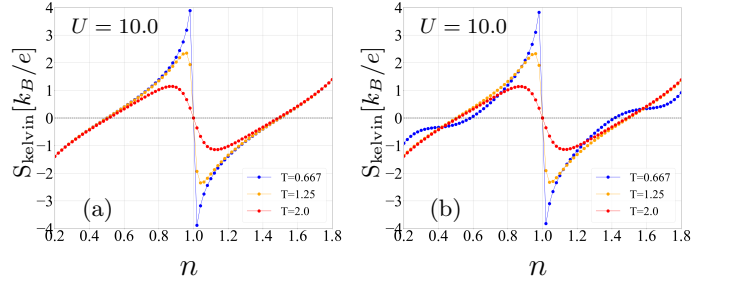

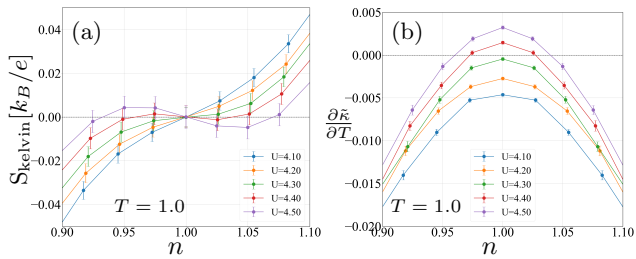
\begin{figure}[t]
\begin{tikzpicture}
\node (img1) {\includegraphics[width=0.45\linewidth]{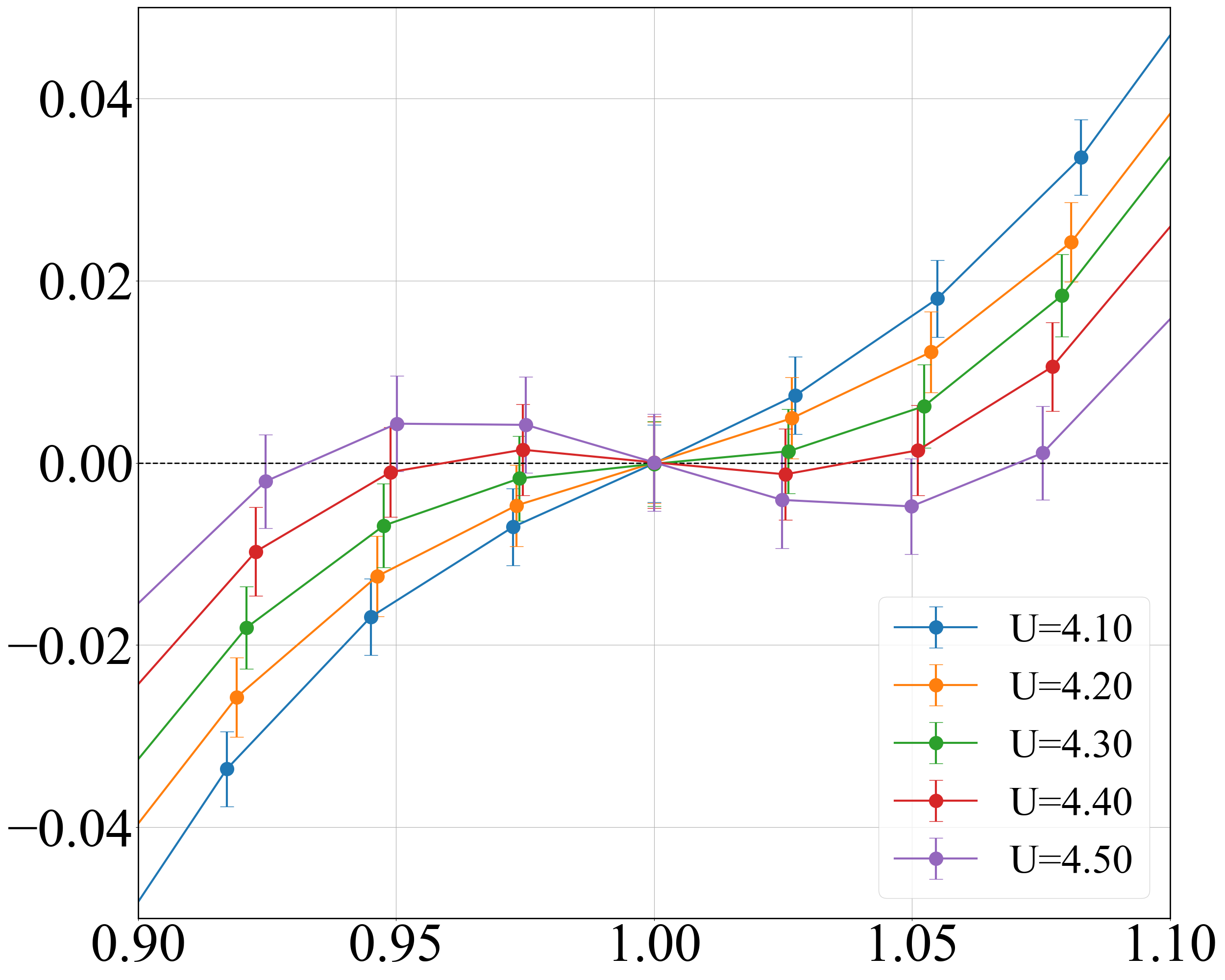}};
\node[left=of img1,node distance=0cm,rotate=90,anchor=center,yshift=-1.0cm,xshift=0.0cm]{\small{ S$_{\text{kelvin}}[k_B/e]$}};
\node[left=of img1,node distance=0cm,yshift=-1.1cm,xshift=3.3cm]{\scriptsize{$T=1.0$}};
\node[below=of img1,node distance=0cm,yshift=1.2cm,xshift=0.0cm]{\normalsize{$n$}};
\node[left=of img1,node distance=0cm,yshift=1.2cm,xshift=2.2cm]{\small{(a)}};
\node (img2) [right=of img1,xshift=-0.95cm]{\includegraphics[width=0.45\linewidth]{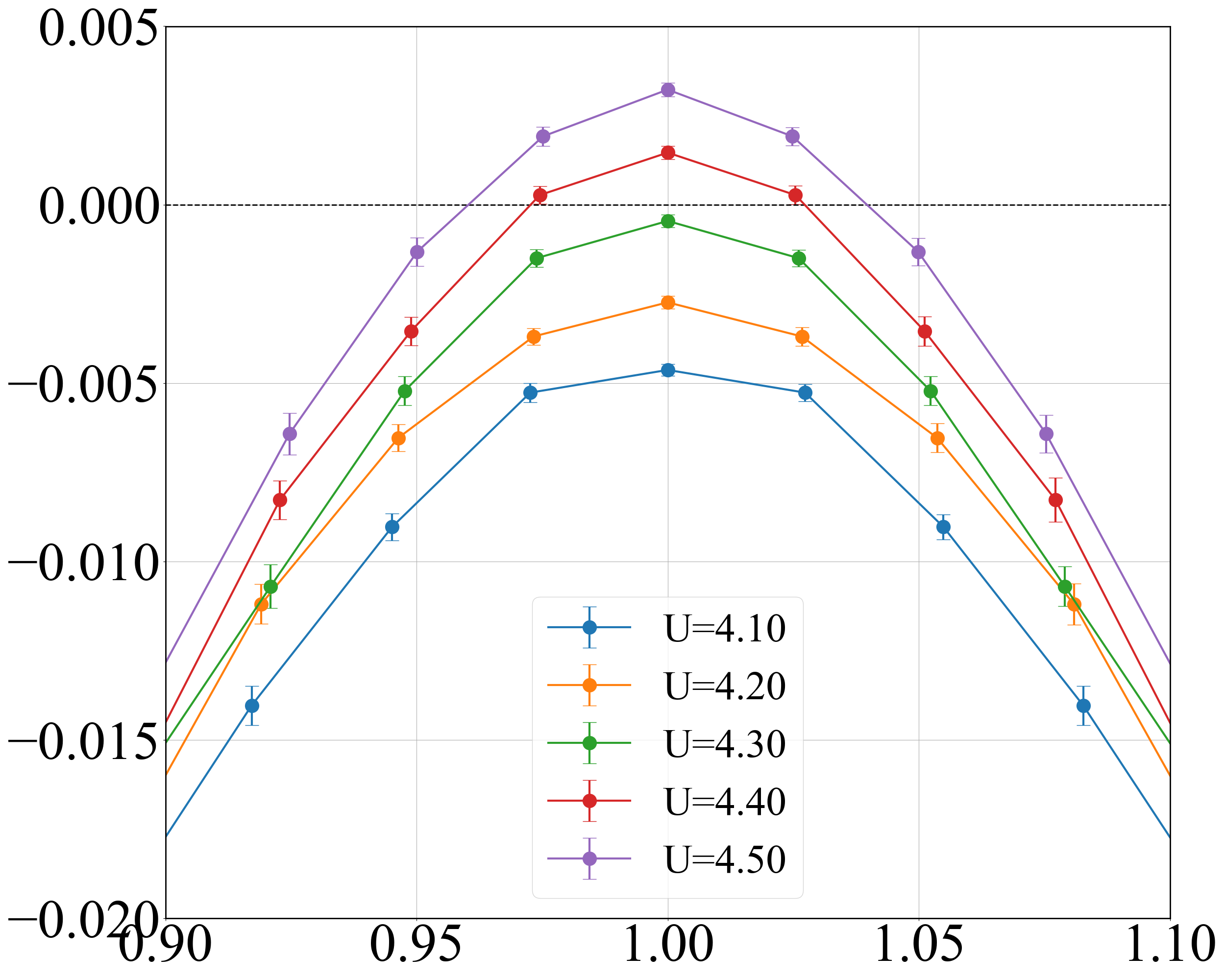}};
\node[left=of img2,node distance=0cm,rotate=0,anchor=center,xshift=1.0cm,yshift=0.0cm]{\small{ $\frac{\partial \tilde{\kappa}}{\partial T}$}};
\node[left=of img2,node distance=0cm,yshift=-1.1cm,xshift=2.8cm]{\scriptsize{$T=1.0$}};
\node[below=of img2,node distance=0cm,yshift=1.2cm,xshift=0.0cm]{\normalsize{$n$}};
\node[left=of img2,node distance=0cm,yshift=1.2cm,xshift=2.3cm]{\small{(b)}};
\end{tikzpicture}
\captionof{figure}{Seebeck coefficient across the metal to insulator crossover. \textbf{(a)} S$_{\rm kelvin}$ at $T = 1.0$. \textbf{(b)} Metal to insulator crossover dictated by $\frac{\partial \tilde{\kappa}}{\partial T}$ at $T=1.0$. Note that an anomalous zero crossing appears only when a charge gap opens up in TDOS at half filling $\frac{\partial \tilde{\kappa}}{\partial T}$.}
\label{Seebeck_across_crossover}
\end{figure}

The parton construction described above is sufficient to give the strong enhancement of the Seebeck coefficient near half filling and its anomalous sign change at intermediate doping, as shown in Fig.~\ref{Seebeck_parton}. The presence of only the density terms, $n_d$, $n_h$ and $n_f$ are sufficient to capture the formation of the Mott plateau in the equation of state curve, due to the presence of $U\hat{d}^\dagger_i\hat{d}_i$ term in the Hamiltonian~\ref{eq:Heff}. As seen in Fig.~\ref{Seebeck_parton}(a), the anomalous sign, as well as the large divergence of the Seebeck coefficient near half filling all follow the formation of the Mott gap in $\tilde{\kappa}=\frac{\partial n}{\partial \mu}$. We also note that the parton theory with only density terms capture the correct temperature dependence of the Seebeck divergence near half filling as found in the QMC simulation [see Fig.~\ref{Seebeck_coefficient}(b)]. However, the anomalous zero crossing at finite doping is ``insensitive" to temperature, and always sits at $n= 0.5, 1.5$. This can be understood by noting that in this limit, entropy is strictly configurational in terms of holon, doublon and spinon occupation on the lattice, which peaks at $n=0.5, 1.5$.

The presence of coherent hopping of doublons, holons and spinons ($\chi$ mean field order parameters in Eq.~\ref{Mean_field_order_parameter}) are enough to generate the temperature variation of $n_s$, as shown in Fig.~\ref{Seebeck_parton}(b). In the parton description, increasing $T$ leads to an increase of $n_s$, in accordance with QMC, however the exact values are different due to thermal fluctuation which QMC captures more accurately than a mean field description. Regardless, it highlights how charge physics is the sole determining factor in the Seebeck anomaly and the temperature dependence of $n_s$ comes from charge excitations moving around. 

\section{Seebeck coefficient across the Metal to Insulator crossover}
\label{Seebeck_MIT}

In Fig.~\ref{Entropy_charge_gap}, we proposed a phase diagram of the Seebeck anomaly through the temperature variation of the thermodynamic density of states (TDOS), $\tilde{\kappa} = \frac{\partial n}{\partial \mu}$ at half filling. To show that this is indeed the case, we take a constant temperature cut across the phase diagram along $T=1.0$. The evolution of the Seebeck coefficient, as one moves to the right in the phase diagram, is shown in Fig.~\ref{Seebeck_MIT}(a), and the evolution of the TDOS through its temperature derivative is shown in (b). At a critical $U_{c}(T)$, there is a crossover from a metallic to insulating behavior at and near half filling due to $\frac{\partial \tilde{\kappa}}{\partial T}>0$. Note that at this $U_c(T)$, an anomalous sign change of the Seebeck coefficient also develops near half filling, showing that the anomalous sign change of the Seebeck coefficient near half filling is driven by a metal to insulator crossover in the TDOS.

\typeout{}
\bibliography{ref}

\end{document}